\documentclass[12pt]{iopart}
\expandafter\let\csname equation*\endcsname\relax
\expandafter\let\csname endequation*\endcsname\relax
\usepackage{amsfonts} 
\usepackage{amsmath}
\usepackage{iopams}
\usepackage{amssymb}
\usepackage{mathtools}
\usepackage{cancel} 
\usepackage{bm}
\usepackage[usenames, dvipsnames]{color} 
\usepackage{dcolumn}
\usepackage{epic} 
\usepackage{epsfig}
\usepackage{feynmp}
\usepackage{graphicx}
\usepackage{grffile}
\usepackage[breaklinks,colorlinks = true,linkcolor = red,urlcolor=blue,citecolor=red]{hyperref}
\usepackage{mathrsfs}
\usepackage{soul}
\usepackage{subfigure}
\usepackage{wrapfig}
\usepackage{xy} 
\usepackage{xcolor}
\usepackage{amsmath,amssymb,amsfonts,amsthm}
\usepackage[ascii]{inputenc}
\usepackage{makecell}
\usepackage{cite}
\allowdisplaybreaks
\usepackage{etoolbox}
\usepackage[normalem]{ulem}

\makeatletter
\newrobustcmd{\fixappendix}{%
  \patchcmd{\l@section}{1.5em}{7em}{}{}%
  \patchcmd{\l@subsection}{2.3em}{7em}{}{}%
}
\makeatother

\graphicspath{{images/}{../Figures}}

\begin{document}
%%%%%%%%%%%%%%%%%%%%%%%%%%%

%%%%%% TITLE %%%%%%%%%%%%%%
\title{Edge fluctuations and third-order phase transition in harmonically confined long-range systems}
%%%%%%%%%%%%%%%%%%%%%%%%%%%
\author{Jitendra Kethepalli, Manas Kulkarni,  Anupam Kundu}
\address{International Centre for Theoretical Sciences, Tata Institute of Fundamental Research, Bengaluru -- 560089, India}
\author{Satya N. Majumdar}
\address{LPTMS, CNRS, Univ.  Paris-Sud,  Universit\'e Paris-Saclay,  91405 Orsay,  France}
\author{David Mukamel}
\address{Department of Physics of Complex Systems, Weizmann Institute of Science, Rehovot 7610001, Israel}
\author{Gr\'egory Schehr}
\address{Sorbonne Universit\'e, Laboratoire de Physique Th\'eorique et Hautes Energies, CNRS UMR 7589, 4 Place Jussieu, 75252 Paris Cedex 05, France}
%%%%%%%%%%%%%%%%%%%%%%%%%%%
%%%%%%%% ABSTRACT %%%%%%%%%
%\tableofcontents
\begin{abstract}
We study the distribution of the position of the rightmost particle $x_{\max}$ in a $N$-particle Riesz gas in one dimension confined in a harmonic trap. The particles interact via long-range repulsive potential, of the form  $r^{-k}$ with $-2<k<\infty$ where $r$ is the inter-particle distance. In equilibrium at temperature $O(1)$, the gas settles on a finite length scale $L_N$ that depends on $N$ and $k$. We numerically observe that the typical fluctuation of $y_{\max} = x_{\max}/L_N$ around its mean is of $O(N^{-\eta_k})$. Over this length scale, the distribution of the typical fluctuations has a $N$ independent scaling form. We show that the exponent $\eta_k$ obtained from the Hessian theory predicts the scale of typical fluctuations remarkably well. The distribution of atypical fluctuations to the left and right of the mean $\langle y_{\max} \rangle$ are governed by the left and right large deviation functions, respectively. We compute these large deviation functions explicitly $\forall k>-2$. We also find that these large deviation functions describe a pulled to pushed type phase transition as observed in Dyson's log-gas ($k\to 0$) and $1d$ one component plasma ($k=-1$). Remarkably, we find that the phase transition remains $3^{\rm rd}$ order for the entire regime. Our results demonstrate the striking universality of the $3^{\rm rd}$ order transition even in models that fall outside the paradigm of Coulomb systems and the random matrix theory. We numerically verify our analytical expressions of the large deviation functions via Monte Carlo simulation using an importance sampling algorithm.
\end{abstract}
%\maketitle
\date{\today}
%%%%%%%%%%%%%%%%%%%%%%%%%%%
\maketitle
%\tableofcontents
%%%%%%%%%%%%%%%%%%%%%%%%%%%
\section{Introduction}
\label{intro} 
Understanding the properties of interacting many particle systems has been a subject of immense interest in both physics and mathematics. Examples of such systems range from sand-pile~\cite{Dhar1999TheAS} to neural networks~\cite{Rotskoff2018NeuralNA}, electrons in metal and quantum liquids~\cite{mahan20089} to finance~\cite{Akemann2011TheOH}, Big-data~\cite{he2015big}, charged particles~\cite{brown2003rotational} and gravitational systems~\cite{padmanabhan1990gravity} to name a few. While collective phenomena are widely studied in many of these systems, recently there has been a growing interest in investigating the local properties such as fluctuations, correlations and extreme value statistics (EVS). With recent developments in experimental techniques it has become possible to probe the physics at a microscopic scale such as in cold atoms~\cite{bakr2009quantum, cheuk2015quantum, parsons2015site} and ions~\cite{zhang2017observation}. Often the physics becomes even more interesting and exotic when the interactions become long-ranged in such systems. Therefore, there is a growing need to study the properties of long range interacting systems.

A suitable and promising platform for such a study is the family of confined Riesz gas models. This consists of $N$ particles interacting via a power law interaction $V(r) \sim 1/r^{k}$, where $r$ is the paiwise inter-particle distance, with $-2<k<\infty$ and confined in a harmonic potential. The (potential) energy function  of the gas is~\cite{marcelriesz1938}
\begin{equation}\label{microhami}
    E_k(\{x_i\}) = \sum_i^N \frac{x_i^2}{2} + \frac{J {\rm sgn}(k)}{2}\sum_{i \neq j}^N |x_i-x_j|^{-k},  
\end{equation}
where $x_i$ is the position of $i^{\rm th}$ particle, $J>0$ is the strength of interaction, $k$ is the exponent of interaction and the sign function $\rm sgn(k)$ ensures the repulsive nature of the interaction. In thermal equilibrium at inverse temperature $\beta$ (assuming Boltzmann constant $k_B =1$), the probability distribution of the positions is given by
\begin{equation}
P^{\rm joint}_{k}\left[\{x_i\}\right] = \frac{e^{-\beta E_k(\{x_i\})}}{Z_k\left(\beta, N\right)}
\end{equation}
where $Z_k\left(\beta, N\right)$ is the partition function that normalizes this probability distribution.

 It has been shown that, in the large $N$ limit, the density of these particles in thermal equilibrium has a finite support~\cite{agarwal2019harmonically}. In this paper, we study the fluctuations of the position $x_{\max}$ of the rightmost particle. This question falls under the paradigm of EVS of correlated variables~\cite{evs_review}. Such questions have been studied in several contexts, for example, random matrix theory (RMT)~\cite{majumdar2014top, nadal2011simple, tracy1994fredholm, tracy1996orthogonal}, the lowest energy modes in ultra cold gas~\cite{giorgini2008theory}, highest energy barrier in disordered systems~\cite{le2003exact}, height fluctuations in interface problems~\cite{majumdar2004exact, schehr2006universal, raychaudhuri2001maximal} and in binary search problems~\cite{krapivsky2000traveling} to name a few.

\begin{figure}[t]
    \centering
    \includegraphics[scale=0.2]{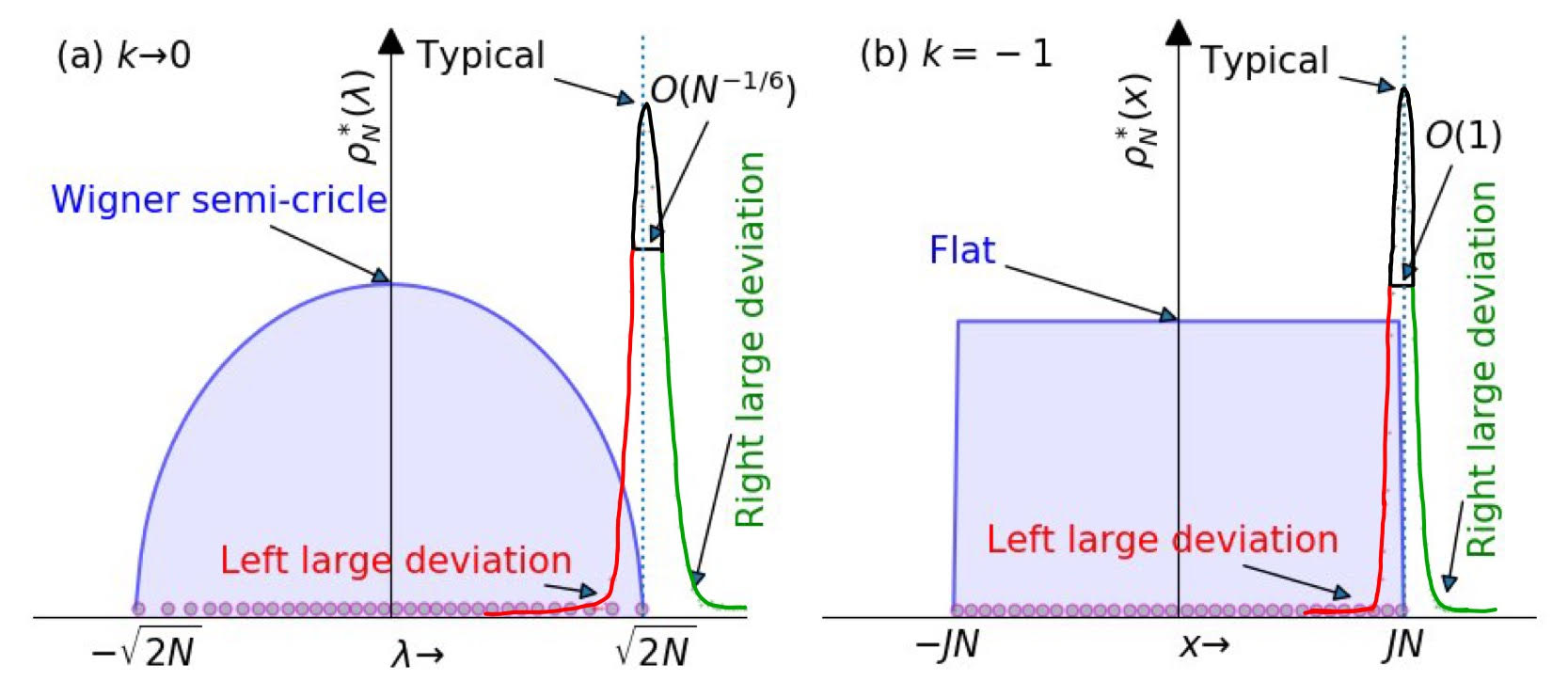}
    \caption{Schematic plot of the probability density function (PDF)  of (a) the largest eigenvalue in RMT (Dyson's log-gas) and (b) the position of the rightmost particle in the 1dOCP, along with the respective density profiles. The PDF of the position of the edge particle in these cases are divided into three parts -- typical (black) in the central part and, left (red) and right (green) large deviations. We show in this paper that such representative pictures also hold for the harmonically confined Riesz gas with $-2<k<\infty$.}
    \label{fig:schematic}
\end{figure}

In the context of RMT, the position $x_{\max}$ of the rightmost particle corresponds to the largest eigenvalue $\lambda_{\max}$ of a $N \times N$ random matrix. For random matrices chosen from Gaussian ensembles characterised by the symmetry class parameter $\beta = 1,2,4$, the joint distribution of the real eigenvalues $\{\lambda_1, \lambda_2....\lambda_N\}$ is given by~\cite{mehta2004random, forrester2010log, majumdar2014top, dean2008extreme} 
\begin{equation}\label{rmt}
    P^{\rm joint}_{0}[\{\lambda_i\}] = \frac{1}{Z_0\left(\beta, N\right)}e^{-\frac{\beta}{2}\left( \sum_{i} \lambda_i^2 - \sum_{i \neq j} \ln|\lambda_i-\lambda_j|\right)},
\end{equation}
where $Z_0\left(\beta, N\right)$ is a normalization constant. This distribution can be interpreted as the Boltzmann weight of $N$ particles with positions $x_i \equiv \lambda_i$ interacting via logarithmic potential. This system of particles is known in the literature as the Dyson's log-gas~\cite{forrester2010log}.  Note that this system corresponds to taking $k \to 0$ limit of the Riesz gas [Eq.~\eqref{microhami}] after the substitution $J \to J_0/k$. Hence, we use subscript $``0"$ in Eq.~\eqref{rmt} and set $J_0=1$. It is well known that for large $N$ the particle (or eigenvalue) density is given by Wigner semi-circle law i.e.
\begin{equation}
\rho^*_N(\lambda) = \frac{1}{\sqrt{N}} f_0\left(\frac{\lambda}{\sqrt{N}}\right) \text{  with  } f_0(y) = \frac{1}{\pi} \sqrt{2-y^2}.
\end{equation}
with the support $\lambda \in [-\sqrt{2 N}, \sqrt{2N}]$~\cite{dyson1962statistical1, dyson1962statistical2}. The largest eigenvalue $\lambda_{\max} = \max_{1 \leq i\leq N}\{\lambda_i\}$ represents the position of the rightmost particle of the log-gas. The statistics of $\lambda_{\max}$ is well understood~\cite{majumdar2014top, dean2008extreme, dean2006large, tracy1994fredholm, tracy1996orthogonal}. In particular, the average of $\lambda_{\max}$ is given by the upper edge of the Wigner semi circle $\langle \lambda_{\max} \rangle = \sqrt{2N}$. The typical fluctuations around this mean are known to scale as $\sigma_{\lambda_{\max}} = \sqrt{\langle \lambda^2_{\max} \rangle - \langle \lambda_{\max} \rangle^2 }  \sim N^{-\frac{1}{6}}$ and are described by the Tracy-Widom distribution ${\cal F}'_\beta(y) = {\cal F}^{(0)'}_\beta$, where the superscript `(0)' refers to the limit $k \to 0$~\cite{tracy1994fredholm, tracy1996orthogonal}. The distribution of atypically large fluctuations of $\lambda_{\max}$ of $O(\sqrt{N})$ on the both sides of the mean (left and right) are described by appropriate large deviation functions (LDF). A schematic plot of this distribution is shown in Fig.~\ref{fig:schematic}a. The cumulative distribution function (CDF) of the scaled variable $\tilde{\lambda}_{\max} = \lambda_{\max}/\sqrt{N}$ is given by~\cite{tracy1994fredholm, tracy1996orthogonal, dean2008extreme, dean2006large, majumdar2009large, majumdar2014top}
\begin{equation}\label{k0CDF}
    {\rm Prob.}[\tilde{\lambda}_{\max}<w, N] \approx
    \begin{cases}
        e^{-\beta N^2 \Phi_{-}(w, 0)} & \sqrt{2}-w\sim O(1) \\
       \mathcal{F}^{(0)}_{\beta} \left(\sqrt{2} N^{\frac{2}{3}}\left(w-\sqrt{2}\right)\right)& |\sqrt{2}-w|\sim O(N^{-\frac{2}{3}}) \\
      e^{-\beta N \Phi_{+}(w, 0)} & w - \sqrt{2}\sim O(1),
    \end{cases}
\end{equation}
where $\Phi_{-}(w, 0)$ and  $\Phi_{+}(w, 0)$ are, respectively, the left and the right LDF. %$\mathcal{F}_{\beta}(x)$ is Tracy-Widom function, for $\beta = 1,2,4$, originally derived in~\cite{tracy1994fredholm, tracy1996orthogonal}.
The $0$ in the argument of the LDF indicates that the log-gas corresponds to Riesz gas in Eq.~\eqref{microhami} in the limit $k \to 0$. These functions have been explicitly computed and are given by~\cite{dean2006large, dean2008extreme} 
\begin{equation}\label{left0}
\Phi_-(w, 0) = \frac{1}{108} \left[36 w^2 -w^4 -(15w+w^3)\sqrt{w^2+6}+27\left(\ln{18}-2\ln[{w+\sqrt{w^2+6}}]\right)\right],
\end{equation} 
for $w<\sqrt{2}$ and~\cite{majumdar2009large, arous2001aging} 
\begin{equation}\label{right0}
\Phi_+(w, 0) = \frac{1}{2}w\sqrt{w^2-2}+\ln{\frac{w-\sqrt{w^2-2}}{\sqrt{2}}}, \text{  for  } w>\sqrt{2}.
\end{equation}
The large deviation behaviour is different for $w>\sqrt{2}$ and $w<\sqrt{2}$. This difference gets manifested as a thermodynamic phase transition if one considers the free energy density given by
\begin{equation}
 \lim_{N \to \infty} -\frac{1}{N^2}\log\left({\rm Prob.}[\tilde{\lambda}_{\max}<w, N]\right) = 
 \begin{cases}
  \Phi_{-}(w, 0), & w<\sqrt{2}\\
  0 & w>\sqrt{2}.
  \end{cases}
\end{equation}
Since $\Phi_-(w, 0) \sim \left(\sqrt{2}-w\right)^{3}$ as $w \to \sqrt{2}$ from Eq.~\eqref{left0}, the third derivative of the free energy with respect to $w$ is discontinuous at $w=\sqrt{2}$. This implies that the system undergoes a $3^{\rm rd}$ order phase transition, from a phase ($w > \sqrt{2}$) in which the rightmost particle is pulled out from the bulk to the right of $w = \sqrt{2}$ (pulled phase) to a phase ($w < \sqrt{2}$) in which all the particles are pushed to the left of $w=\sqrt{2}$ (pushed phase)~\cite{majumdar2014top}. Recall that $w= \sqrt{2}$ is the right edge of the scaled density of the particles.
%the pushed phase with $w<\sqrt{2}$ (all the particles are pushed to the left) and free energy $N^2\Phi_-(w, 0)$ to the pulled phase with $w>\sqrt{2}$ (rightmost particle is pulled from the bulk $\in \left[-\sqrt{2}, \sqrt{2}\right]$) and free energy $N\Phi_+(w, 0)$.

A similar transition has also been observed in the $1d$ one component plasma (1dOCP) confined by a harmonic potential. The energy function in this case is given by
\begin{equation}\label{microhami-1}
    E_k(\{x_i\}) = \sum_i^N \frac{x_i^2}{2} - \frac{J}{2}\sum_{i \neq j}^N |x_i-x_j|,  
\end{equation}
where $x_i$'s are the particle positions and $J$ is the strength of the repulsive interaction. Note that this corresponds to $k=-1$ of the Riesz gas model Eq.~\eqref{microhami}. Here the average thermal density profile is flat and is given by
\begin{equation}\label{wflt}
\rho_{N}^*(x) = \frac{1}{N}f_{-1}\left(\frac{x}{N}\right)  \text{  where  } f_{-1}(y) = \frac{1}{2 J},
\end{equation}
with the support $x \in [-N J , N J]$. The statistics of the position of the rightmost particle $x_{\max}$ has been studied recently~\cite{dhar2017exact, dhar2018extreme, rojas2018universal,flack2021truncated}. Its average is $\langle x_{\max} \rangle = N J$ and its typical fluctuations are $O(1)$ and are governed by the CDF $\mathcal{F}^{(-1)}_{\beta}(x)$ which is a solution to a non-local eigenvalue equation
\begin{equation}\label{typical-ocp}
    \frac{d}{dx} \mathcal{F}^{(-1)}_{\beta}\left( x \right)  = A\left(J\right) e^{-\frac{x^2}{2}}\mathcal{F}^{(-1)}_{\beta}\left( x +2 J\right),
\end{equation}
where the eigenvalue $A(J)$ is determined by satisfying the boundary conditions $ \mathcal{F}^{(-1)}_{\beta}(-\infty) = 0$, $ \mathcal{F}^{(-1)}_{\beta}(\infty) = 1$ and $0\leq \mathcal{F}^{(-1)}_{\beta}(x) \leq 1$ for $x \in (-\infty, \infty)$. The distribution of atypical fluctuations (of $O(N)$ from the mean) governed by the LDF $\Phi_{\pm}(w,-1)$ are also well understood. A schematic plot of the PDF is shown in Fig.~\ref{fig:schematic}b. The CDF of the scaled variable $y_{\max} = x_{\max}/N$ is given by
\begin{equation}\label{km1CDF}
    {\rm Prob.}[y_{\max}<w, N] \approx
    \begin{cases}
      e^{-\beta N^3 \Phi_{-}(w,-1)} & J-w\sim O(1) \\
      \mathcal{F}^{(-1)}_{\beta}\left(N\left(w -J \right)+J\right)& |w -J |\sim O\left(N^{-1}\right) \\
      e^{-\beta N^2 \Phi_{+}(w, -1)} & w -J\sim O(1).
    \end{cases}
\end{equation}
The LDF are given by~\cite{dhar2017exact, dhar2018extreme} 
\begin{align}\label{leftm1}
\Phi_-(w, -1)  &= \begin{cases}
	\frac{w^2}{2} + \frac{J^2}{6} & \text{  for  } \quad  w<-J\\
	\frac{(J-w)^3}{12 J} & \text{  for  } -J< w < J,
\end{cases}\\
\label{rightm1}
\Phi_+(w, -1) &= \frac{(w-J)^2}{2}.
\end{align}
Analogous to the log-gas case [Eq.~\eqref{k0CDF}], for the 1dOCP also the large deviation functions exhibit different behaviours to the left and right of the mean position $\langle y_{\max} \rangle= J$.  Once again this difference gets manifested as a pulled to pushed phase transition at $w=J$. Interestingly the order of the phase transition is also $3$ because $\Phi_-(w, -1) \sim \left(J-w\right)^{3}$ as $w \to J$ from Eq.~\eqref{leftm1}~\cite{dhar2018extreme}.

There are many physical problems where these pulled to pushed type of phase transitions have been investigated. For example, such transitions has been observed in spin-glass~\cite{fyodorov2012critical}, wireless telecommunication~\cite{kazakopoulos2011living}, chaotic cavities~\cite{vivo2008distributions, vivo2010probability, damle2011phase, cunden2015joint}, entanglement in bipartite quantum systems~\cite{de2010phase, facchi2008phase, nadal2010phase}, random tilings~\cite{colomo2013third} and non intersecting Brownian excursions~\cite{schehr2013reunion, forrester2011non} to name a few (a review can be found in Ref.~\cite{majumdar2014top}). Since these systems are often related to RMT, the third-order transition is attributed to Dyson's log-gas and its variants. Another family of models different from log-gas which also exhibit such third-order phase transitions, are confined particles in $d$ dimensions interacting via $d$ dimensional Coulomb interaction potentials ($V(r)$ is $|r|, \log(r)$ for $d=1,2$ respectively and $V(r) = 1/r^{d-2}$ for $d > 2$)~\cite{cunden2017universality, cunden2018universality} and Yukawa potentials~\cite{cunden2019third}. In fact similar phase transitions were already identified in the context of large-$N$ gauge theories and are well known as Gross-Witten-Wadia~\cite{gross1980possible,  wadia1980n} or Douglas-Kazakov~\cite{douglas1993large} phase transitions.

The third-order phase transitions in all the above studies are either rooted in RMT or Coulomb interaction. In this paper, we investigate the extent of this universality in models which do not fall in either of the above two classes and focus on the Riesz gas family of models which has repulsive interactions of the form $V(r) \propto |r|^{-k}$.

We study the large-deviation properties of the distribution of the position of the rightmost particle of the harmonically confined Riesz gas model with general $k>-2$. We obtain the explicit expressions for the left and the right LDF $\Phi_{-}(w, k)$ and $\Phi_{+}(w, k)$, respectively. We find that for these models also the properties of large deviations get manifested as a pulled to pushed phase transition. Remarkably, we show that the third-order phase transition persists  $\forall k>-2$, thereby demonstrating the universality even beyond RMT and Coulomb class of models. We also study the system size scaling of the typical fluctuations numerically and we find that the commonly used ``Lifshitz argument" is valid only for special values $k=-1$ and $k \to 0$. In addition we also show that the appropriate Hessian theory predicts the scale of the typical fluctuations remarkably well.

The rest of the paper is organised as follows. In Section.~\ref{recap}, we discuss the relevant properties of the Riesz gas and establish some important notations. In Section.~\ref{summary} we provide a summary of our results. The derivation of results are given in  Section.~\ref{distofxmax}. We conclude our findings along with an outlook in Section.~\ref{conclusion}. Additional details of our analytical and numerical results are relegated to~\ref{apndtyp},~\ref{apndixa} and~\ref{apndims}.

\section{Some properties of Riesz gas and important notations}
\label{recap}
In this section, we recap some properties of Riesz gas that are relevant to our studies. We start by noting that the Riesz gas family for different value of $k$ corresponds to various well known models. Some of the well studied cases include $k=2$ which is an interacting integrable system, known as the Calogero-Moser model~\cite{polychronakos2006physics, calogero1971solution, calogero1975exactly,  agarwal2019some, agarwal2019harmonically} and $k =-1$, which corresponds to the one dimensional one component plasma ($1d$OCP)~\cite{baxter1993statistical, dhar2017exact, dhar2018extreme} discussed above. It is interesting to note that the $k \to \infty$ limit, corresponds to the hard-rod gas~\cite{saff1997distributing, kethepalli2021harmonically}. As described earlier the Dyson's log-gas is also a part of Riesz family with the interaction strength $J \to J_0/k$ and $k\to0$. In this paper, we recall that we have used $J_0 = 1$.

Since for $k<-2$ the repulsive interaction term $|x_i-x_j|^{-k}$ is stronger than the confining external potential $x^2$, the particles fly away to $\pm \infty$. Hence we confine our study to $-2<k<\infty$. In the temperature regime $\beta \sim O(1)$ considered here, the repulsive interaction and the confining potential compete with each other and as a result the particles settle down over a finite support of length of order $ O(2 L_N)$ for large $N$. The $N$ dependence of $L_N$ can be estimated by balancing the repulsive interaction term and the confining potential energy to get~\cite{agarwal2019harmonically, kethepalli2021harmonically} 
\begin{equation}\label{scaling}
L_N =
\begin{cases}
N^{\alpha_k} & \text{ for } k \neq 1\\
\left(N \log(N)\right)^{1/3} & \text{ for } k = 1
\end{cases}
\quad \text{with} \quad
\alpha_k = 
\begin{cases}
\frac{1}{k+2} & \text{ for } -2<k<1\\
\frac{k}{k+2} & \text{ for } k>1
\end{cases}
.
\end{equation}
It is thus convenient to scale the positions of the particles by this length scale $L_N$ and set $y_i = x_i/L_N$. We use these scaling variables to present the results and discussions. In a recent study~\cite{agarwal2019harmonically}, the large-$N$ field theory of the Riesz gas model has been established. The saddle point calculation in the field theory shows that the average density profile is given by $\rho_N^*(x) = \frac{1}{L_N} \rho_{k, {\rm uc}}^{\rm{ *}}\left(\frac{x}{L_N}\right)$, with the scaling form (with $y = x/L_N$ being the scaled variable)
\begin{equation} \label{rho_uc}
    \rho_{k, {\rm uc}}^{\rm{*}}(y) = \frac{\left(1-(y/l_k^{\rm uc})^2\right)^{\gamma_k}}{2^{{2\gamma_k+1}} l_k^{\rm uc}B\left(\gamma_k+1, \gamma_k+1\right)},~~\text{for}~ -l_k^{\rm uc} \leq y \leq l_k^{\rm uc},
\end{equation}
where the subscript `uc' refers to the unconstrained Riesz gas and 
\begin{equation}
l_k^{\rm uc} = \frac{1}{2} \left(A_k B\left(\gamma_k+1, \gamma_k+1\right)\right)^{-\alpha_k} 
~ \text{with} ~
A_k = 
\begin{cases}
\left(2(k+1) \zeta(k)\right)^{-\gamma_k}, & \text{ for } k>1\\
\frac{1}{4} & \text{ for } k=1\\
\frac{\sin[\pi \gamma_k]}{2 \pi \gamma_k |k|} & \text{ for } -2<k<1.
\end{cases}
\label{A_1k}
\end{equation}
Here $B(x,y)$ is the standard Beta function, $\zeta(k) = \sum_{n=1}^{\infty} 1/n^{k}$ is the Riemann zeta function and 
\begin{equation}
\gamma_k =  
\begin{cases}
\frac{k+1}{2} & \text{ for } -2<k<1\\
\frac{1}{k} & \text{ for } k>1
\end{cases}
.
\label{gammak}
\end{equation}
Note that Eq.~\eqref{A_1k} is valid for $k \neq 0$. However for $k \to 0$, as mentioned previously, one has to substitute $J\to 1/k$ to get a sensible limit. In this limit, we find $l_0^{\rm uc} = \sqrt{2}$ and $A_0 = 1/2\pi$. The form of this density profile is different for different values of $k$. Depending on the behaviour of this profile at the edge, the range $-2<k<\infty$ can be organised into three regimes~\cite{kethepalli2021harmonically}: (1) $k>1$ (short-ranged), (2) $-1<k<1$ (weakly long-ranged) and (3) $-2<k \leq -1$ (strongly long-ranged). In this paper, we have set $J=1$ for $k \neq 0$ while, for $k \to 0$, we take $J \to 1/k$.

In the presence of a hard wall (on the right of the gas), these density profiles get modified differently  for different values of $k$. This problem was studied recently  in Ref.~\cite{kethepalli2021harmonically} and the results of this study lay the foundation of the current work (as will be elaborated later). In the next section we summarize our main findings.

\section{Summary of the results}
\label{summary}

\begin{figure}[t]
    \centering
    \includegraphics[scale=0.8]{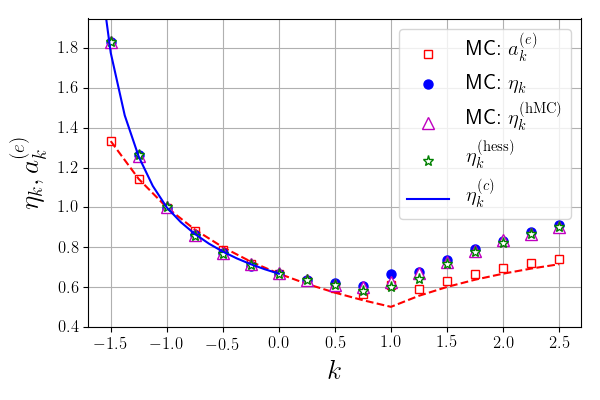}
    \caption{Behaviour of $\eta_k$ [Eq.~\eqref{eta-k}, disks] and the exponent $a_k^{(e)}$ of the mean gap at the edge [Eq.~\eqref{lifshitz}, squares], as a function of $k$ is plotted here. The exponent $\eta_k$ for different values of $k$ are obtained by fitting the data for $\sigma^2_{y_{\max}}$ with $N$ [See Fig.~\ref{fig:etakall} and~\ref{apndtyp}] obtained numerically (MC) for the confined Riesz gas [Eq.~\eqref{microhami}]. The exponent $a_k^{(e)}$ is obtained similarly. We notice that $a_k^{(e)}$ obtained from numerics agrees with the one obtained from the Lifshitz argument (dashed red line) given in Eq.~\eqref{lifshitz-arg}. Green stars represent the exponent $\eta^{\rm hess}_k$ obtained by numerically inverting the Hessian matrix [see Eq.~\eqref{hess1} and Eq.~\eqref{hess2}]. We also find the exponent $\eta_k^{\rm (hMC)}$ [See Fig.~\ref{fig:etakall_h} and~\ref{apndtyp}] from the MC simulations of the Hessian hamiltonian [Eq.~\eqref{hess:hami}]. It is interesting to note that the value of the exponent extracted from the three different approaches are in excellent agreement with each other ($\eta_k = \eta_k^{\rm (hess)} = \eta_k^{\rm (hMC)}$). The solid blue line represents the conjecture for $\eta_k$ for $-2<k<0$ given in Eq.~\eqref{conjecture} (where the superscript ``$c$" indicates our conjecture). The agreement between this conjecture and the MC simulation results is excellent. The parameters used in these simulations are $T=1$ and $J=1$.}
    \label{fig:etakvsk}
\end{figure}

\begin{figure}[t]
    \centering
    \includegraphics[scale=0.4]{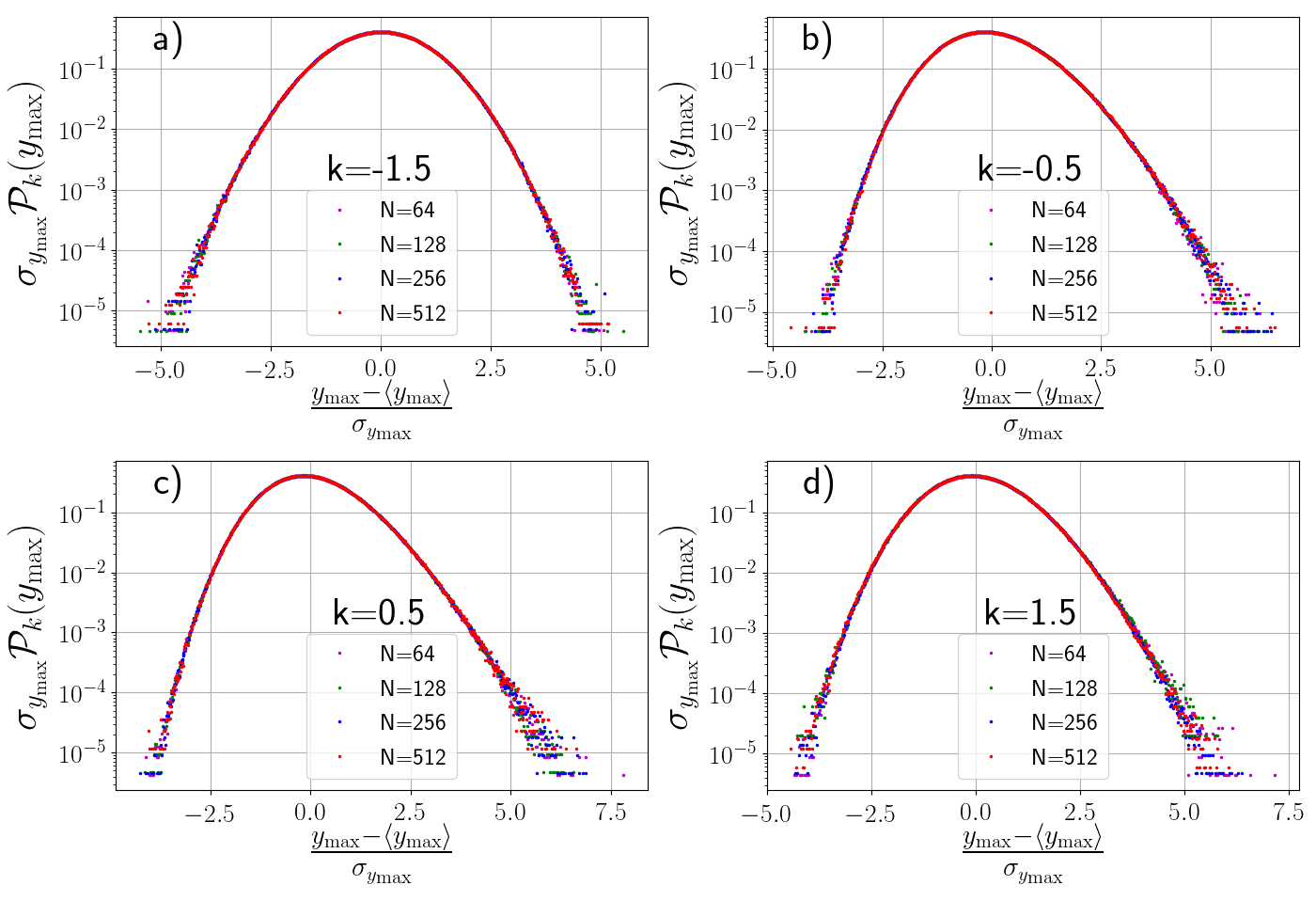}
    \caption{Typical distribution of $y_{\max}$ for various values of $k$. We notice excellent data collapse, when we plotted $\sigma_{y_{\max}} \mathcal{P}_k(y_{\max}, N)$ versus the scaling variable $(y_{\max} - \langle y_{\max} \rangle)/\sigma_{y_{\max}}$. Here the values of $\sigma_{y_{\max}}$ and $\langle y_{\max} \rangle$ were extracted from the data (therefore no fitting parameter was used).}
    \label{fig:distr}
\end{figure}
\begin{figure}[t]
    \centering
    \includegraphics[scale=0.5]{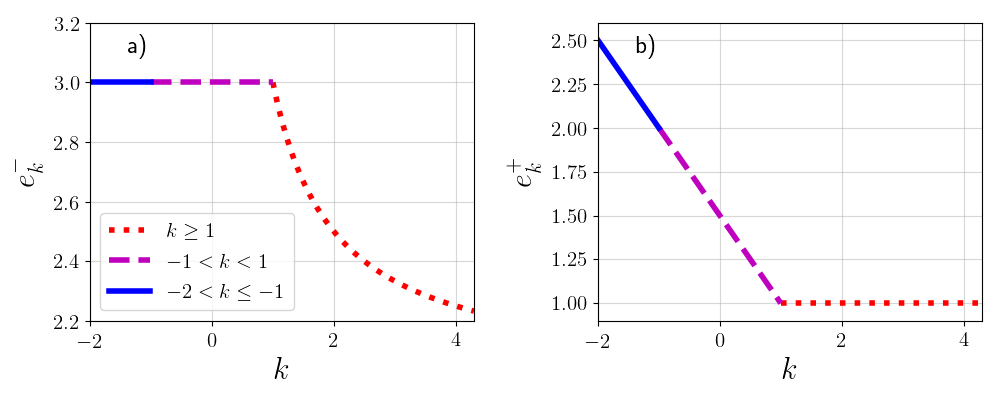}
    \caption{(a) Plot of the exponent $e_k^{-}$, that governs the asymptotic behaviour of the left large deviation function  given in Eq.~\eqref{phimekmkg1} for $k \geq 1$, Eq.~\eqref{asym:ldf} for $-1<k<1$ and Eq.~\eqref{phimekmklm1} for $-2<k \leq -1$. 
    (b) Plot of the exponent $e_k^+$ characterising the asymptotic behaviour of the right LDF which is $1$ for $k \geq 1$, and is given in Eq.~\eqref{asym:rdf} for $-1<k<1$ and Eq.~\eqref{phipekpklm1} for $-2<k \leq -1$.}
    \label{fig:ekg1}
\end{figure}

We study the distribution of the typical and the atypical fluctuations of the scaled position $y_{\max}$ of the rightmost particle for different $k$. The typical part of the distribution is studied numerically and the atypical part is studied both analytically and numerically. We show that the atypical fluctuations are described by the appropriate LDF. The mean of the rightmost particle is given by the upper edge of the support of unconstrained density [Eq.~\eqref{A_1k}], which in scaled variable $y_{\max} = x_{\max}/L_N$ is given by $\langle y_{\max} \rangle = l_k^{\rm uc}$. For large but finite $N$, $y_{\max}$ fluctuates from sample to sample and we numerically observe that the standard deviation $\sigma_{y_{\max}} = \sqrt{\langle y_{\max}^2 \rangle- \langle y_{\max} \rangle^2}$ describing the typical fluctuation is of order $N^{-\eta_k}$. It is known that, for inverse temeprature $\beta = O(1)$, for the Dyson's log-gas $\eta_0 = 2/3$~\cite{majumdar2014top,forrester2010log}, for the 1dOCP $\eta_{-1} = 1$~\cite{dhar2017exact, dhar2018extreme} while  for the Calegoro-Moser system $\eta_2 = 5/6$ \footnote{The values of $\eta_0$~\cite{majumdar2014top,forrester2010log} and $\eta_{-1}$~\cite{dhar2017exact, dhar2018extreme} are analytically established whereas that of $\eta_2$ is numerically established ~\cite{agarwal2019some}.}~\cite{agarwal2019some}. We have computed $\eta_k$ numerically for different values of $k$ via Monte-Carlo (MC) simulation using the Metropolis-Hastings algorithm and the result are shown in Fig.~\ref{fig:etakvsk}. By expanding the energy around the ground state and truncating it at bilinear order, as is done within the Hessian theory, we find that the resulting exponent of the variance fits the numerically obtained exponent remarkably well. Further, we provide a conjecture [Eq.~\eqref{conjecture}] for the explicit $k$ dependence of $\eta_k$ for  $k<0$ based on scaling arguments. Fig.~\ref{fig:etakvsk} demonstrates an excellent agreement between our conjecture and numerical data. We observe an excellent collapse for the typical part of the distribution in terms of the scaling variable $\tilde{y}_{\max} = (y_{\max} - \langle y_{\max} \rangle)/\sigma_{y_{\max}}$ (see Fig.~\ref{fig:distr}). This implies that the CDF has scaling form ${\rm Prob.}[y_{\max}<w, N] = \mathcal{F}^{(k)}_{\beta}\left(N^{\eta_k} \left(w-l_k^{\rm uc}\right)\right)$ for large $N$ in the typical part of the distribution i.e. $ |w-l_k^{\rm uc}| \lesssim O(N^{-\eta_k}) $. The fluctuations larger than this scale i.e. $|w-l_k^{\rm uc}| \sim O(1)$ are atypical fluctuations which are described by the left and the right LDF $\Phi_{-}(w, k)$ and  $\Phi_{+}(w, k)$ respectively. The CDF has the form
\begin{equation}\label{summary:cdf}
{\rm Prob.}\left[y_{\max}<w, N\right] \approx
    \begin{cases}
        e^{-\beta N^{2 \alpha_k +1} \Phi_{-}(w, k) }    & l_k^{\rm uc}-w \gtrsim O(1)\\
        \mathcal{F}^{(k)}_{\beta} \left(N^{\eta_k} \left(w-l_k^{\rm uc}\right)\right) & |w-l_k^{\rm uc}| \lesssim O(N^{-\eta_k})\\
        1-e^{-\beta N^{2 \alpha_k} \Phi_{+}(w, k) } & w-l_k^{\rm uc} \gtrsim O(1),
    \end{cases}
\end{equation}
where $\alpha_k$ is given in Eq.~\eqref{scaling} and equivalently the PDF is given by
\begin{equation}\label{summary:pdf}
\mathcal{P}_k\left[y_{\max}=w, N\right] \approx
    \begin{cases}
        e^{-\beta N^{2 \alpha_k +1} \Phi_{-}(w,k) }    & l_k^{\rm uc}-w \gtrsim O(1)\\
        N^{\eta_k}\mathcal{F}^{(k)'}_{\beta} \left(N^{\eta_k} \left(w-l_k^{\rm uc}\right)\right) & |w- l_k^{\rm uc}| \lesssim O(N^{-\eta_k})\\
        e^{-\beta N^{2 \alpha_k} \Phi_{+}(w,k) } & w-l_k^{\rm uc} \gtrsim O(1).
    \end{cases}
\end{equation}
It is worth reminding that for $k \to 0$, $\mathcal{F}_{\beta}^{(0)'}(z)$ is the Tracy-Widom distribution while for $k=-1$, $\mathcal{F}_{\beta}^{(-1)}(z)$ is the solution of Eq.~\eqref{typical-ocp}. As mentioned above for general $k$, we provide numerical evidence supporting the existence of the scaling distributions $\mathcal{F}^{(k)}_{\beta}(z)$ for other values of $k$ (see Fig.~\ref{fig:distr}).

\begin{table}
\begin{center}
\begin{tabular}{ |c|c|c|c|c| } 
 \hline
 Regimes & $e^{-}_k$ & $e^{+}_k$ & $\Phi_{-}(w,k)$ & $\Phi_{+}(w,k)$ \\ 
 \hline
 \hline
 $k>1$ & $2+\frac{1}{k}$ & $1$ &Eq.~\eqref{ldf:lkg1_1}, Fig.~\ref{fig:k2p5}a& Eq.~\eqref{ldf:rkg1_1}, Fig.~\ref{fig:k2p5}b\\ 
 \hline
 $-1<k<1$ & $3$ & $\frac{3-k}{2}$ &Eq.~\eqref{ldf:lkgm1_2}, Fig.~\ref{fig:k0p5}a& Eq.~\eqref{ldf:rkgm1_3}, Fig.~\ref{fig:k0p5}b\\ 
 \hline 
 $-2<k<-1$ & $3$ & $\frac{3-k}{2}$ &Eq.~\eqref{ldf:lklm1_2}, Fig.~\ref{fig:km1p5}a& Eq.~\eqref{ldf:rklm1_3}, Fig.~\ref{fig:km1p5}b\\ 
 \hline
\end{tabular}
\caption{\label{table:epmk} The table summarizes the exponents $e^{\mp}_k$ that characterize the asymptotic behaviour of left and right LDF respectively. The reference to the expressions together with the corresponding plot of the LDF in various regimes of $k$ are also provided.}
\end{center}
\end{table}

 %Similar to the log-gas ($k \to 0$) and $1d$OCP ($k=-1$) the distribution of atypical values of rightmost particle for general $k$, in scaled variable $w=W/N^{\alpha_k}$, is governed by the left large deviation function $\Phi_{-}(w,k)$ when $w<l_k^{\rm uc}$ and by the right large deviation function $\Phi_{+}(w,k)$ when $w>l_k^{\rm uc}$ as can be seen from the Cumulative distribution function (CDF)
We have analytically studied the probability of the atypical fluctuations characterised by the LDF $\Phi_{\pm}(w, k)$. We have obtained explicit expressions of these functions given in Eq.~\eqref{ldf:lkg1_1} and~\eqref{ldf:rkg1_1} for $k \geq 1$, Eq.~\eqref{ldf:lkgm1_2} and~\eqref{ldf:rkgm1_3} for $-1<k<1$ and Eq.~\eqref{ldf:lklm1_2} and~\eqref{ldf:rklm1_3} for $-2<k<-1$. These explicit expressions are one of our main results. 

Another important result of our study is the observation that for general $k$ also, the PDF behaves differently for $w>l_k^{\rm uc}$ and $w<l_k^{\rm uc}$ (as seen in the Dyson's log-gas and the 1dOCP) which again leads to a phase transition at the $w = \l_k^{\rm uc}$. This transition can be seen from the behaviour of the free energy  (discussed later in the Section.~\ref{dist_aty}), namely
\begin{equation}\label{transition}
\lim_{N \to \infty} -\frac{1}{N^{2\alpha_k+1}}\log\left({\rm Prob.}[y_{\max}<w, N]\right) = 
\begin{cases}
\Phi_{-}(w, k), & w<l_k^{\rm uc}\\
0 & w>l_k^{\rm uc},
\end{cases}
\end{equation}
across $w = l_k^{\rm uc}$.
The nature of the transition is determined by the asymptotic behavior of the left large deviation function as $w \to l_k^{\rm uc-}$, 
\begin{equation}\label{philasym}
 \Phi_{-}(w, k) \sim (l_k^{\rm uc}-w)^{e^-_k},
\end{equation}
where the exponent $e^-_k$ determines the order of the transition. The asymptotic behaviour of the right large deviation function as $w \to l_k^{\rm uc +}$ is given by
\begin{equation}
 \Phi_{+}(w, k) \sim (w-l_k^{\rm uc})^{e^+_k}.
\end{equation} 
In Section.~\ref{distofxmax} we compute the exponents $e^{\mp}_k$, $\forall k>-2$  analytically. The values of these exponents are presented in the Table~\ref{table:epmk} and a representative plot is given in Fig~\ref{fig:ekg1}. In the regime $-2<k<1$, we find that the order of phase transition is $3$, since $e_k^{-}=3$. In the regime $k>1$, $e_k^{-} = (2+1/k)$, which implies that the third derivative of $\Phi_{-}(w, k)$ is discontinuous and hence the system undergoes a third-order phase transition (because $\left \lceil 2+1/k \right \rceil = 3$, for $k>1$ where $\left \lceil.\right \rceil$ represents the ceiling function)  according to the Ehrenfest classification~\cite{goldenfeld2018lectures, stanley1972introduction, sachdev_2011}. This leads to the remarkable finding that $\forall k>-2$ the system exhibits a third-order phase transition \footnote{Alternatively, the Ehrenfest classification~\cite{goldenfeld2018lectures, stanley1972introduction, sachdev_2011} can be generalised by extending the notion of normal derivatives to fractional derivatives~\cite{ma2019fractional, hilfer2000applications, hilfer1992multiscaling} $\frac{d^a}{d w^a} w^b = \frac{\Gamma[b+1]}{\Gamma[b+1-a]} w^{b-a},~\text{with}~a, b>0$. If one goes by this classification the order of phase transition is $\left(2+1/k\right)$ for $k>1$.}.

\section{Distribution of $x_{\max}$}
\label{distofxmax}

We start with the CDF of $x_{\max} = \max_{1\leq i \leq N}{x_i}$, namely 
\begin{equation}\label{CDF:ratio}
\begin{split}
 {\rm Prob.}[x_{\max}<W, N] &= {\rm Prob.}[\{x_i<W\}_{i=1}^N, \beta, N]= \frac{Z_k(W, \beta, N)}{Z_k(W \to \infty, \beta, N)},
\end{split}
\end{equation}
where, the partition function $Z_k(W, \beta, N)$ is given by
\begin{equation}
    Z_k(W, \beta,N) = \int_{-\infty}^{W} dx_1  \ldots \int_{-\infty}^{W} dx_N \, e^{-\beta E_k(\{x_i\})}\,  \;,  \label{Z_N(W)} 
\end{equation}
with $E_k(\{x_i\})$ given in Eq.~\eqref{microhami}. This partition function can be interpreted as the partition function of the original Riesz gas in the presence of a hard wall at $x = W$. For $k \to 0$ and $k =-1$ these multiple integrals can be computed in the large $N$ limit. It has been shown that this integral is related to the solution of Painlev{\'e} equation for $k \to 0$~\cite{tracy1994fredholm, tracy1996orthogonal} and a non-local eigenvalue equation for $k=-1$~\cite{dhar2017exact, dhar2018extreme}. For other values of $k$, performing these multiple integrals analytically remains an open and challenging problem. We therefore resort here to direct numerical simulations to compute the typical part of the distribution.

\subsection{Distribution in the typical region}
\label{dist_ty}

In this section we discuss the distribution of $y_{\max} = x_{\max}/L_N$. To compute this distribution numerically we perform conventional MC  simulations using the Metropolis-Hastings algorithm for different values of $k$ from the three regimes mentioned previously ($k \geq 1$, $-1<k<1$ and $-2< k \leq -1$). For each value of $k$, we perform simulations for $N = 64, 128, 256, 512$ and also compute the $\langle y_{\max} \rangle$ and the variance $\sigma_{y_{\max}}^2 = \langle y_{\max}^2 \rangle-\langle y_{\max} \rangle^2$. As argued before, we expect $\langle y_{\max}\rangle = l_k^{\rm uc}$, for large-$N$, which is indeed corroborated by our simulations. Furthermore, we find that for large $N$, $\sigma_{y_{\max}}$ scales as 
\begin{equation}\label{eta-k}
\sigma_{y_{\max}} \sim N^{-\eta_k},~\text{ with }~ \eta_k>0,
\end{equation}
as shown in Fig.~\ref{fig:etakall} in \ref{apndtyp}. In Fig.~\ref{fig:etakvsk} we plot $\eta_k$ as a function of $k$ where we observe that $\eta_k$ is interestingly non-monotonic. 

One naturally wonders how this fluctuation of $y_{\max}$ compares with the mean of the separation between the scaled positions of the rightmost and the second rightmost particles denoted as $\langle \Delta_{\rm edge} \rangle$. The $N$ dependence of this average separation at the edge can be obtained using the ``Lifshitz argument", which is frequently used in extreme value statistics~\cite{majumdar2014top, evs_review}. According to this argument 
\begin{equation}\label{lifshitz-arg}
N \int_{l_k^{\rm uc}-\langle \Delta_{\rm edge} \rangle}^{l_k^{\rm uc}} dy~\rho^*_{k, \rm uc}(y) = 1, 
\end{equation}
which essentially says that there is only one particle between the positions $l_k^{\rm uc}-\langle \Delta_{\rm edge} \rangle$ and $l_k^{\rm uc}$. This equation implies 
\begin{equation}\label{lifshitz}
\langle \Delta_{\rm edge} \rangle \sim N^{-a_k^{(e)}}~\text{ with } a_k^{(e)} = \frac{1}{1+\gamma_k},
\end{equation}
where $\gamma_k$ is given in Eq.~\eqref{gammak}. Our numerical data (squares) for $\langle \Delta_{\rm edge}\rangle$ verifies this result in Eq.~\eqref{lifshitz} as shown in Fig.~\ref{fig:etakvsk}. It is usually assumed that the average edge gap provides the scale for the fluctuations of $y_{\max}$. This has been confirmed for Dyson's log-gas ($k \to 0$) and the 1dOCP ($k=-1$). Interestingly, our numerical results in Fig.~\ref{fig:etakvsk} show that this assumption $\eta_k = a_k^{(e)}$ is not true  for other values of $k$.

We now look at the distribution of the typical fluctuations (of order $\sim \sigma_{y_{\max}}$) for different values of $k$. In Fig.~\ref{fig:distr} we plot $\sigma_{y_{\max}} P_{\rm num}(y_{\max})$ obtained numerically as functions of $ \frac{y_{\max}-\langle y_{\max} \rangle}{\sigma_{y_{\max}}}$ (where the subscript ``num" represents the distribution obtained from numerics). The excellent data collapse for different values of $N$ indicates the following scaling behaviour for the typical part of the distribution 
\begin{equation}\label{ptyp}
\mathcal{P}_k\left[y_{\max}=w, N\right] \approx N^{\eta_k}\mathcal{F}^{(k)'}_{\beta} \left(N^{\eta_k} \left(w-l_k^{\rm uc}\right)\right), ~\text{for}~|w- l_k^{\rm uc}| \lesssim O(N^{-\eta_k}), 
\end{equation}
as announced in Eq.~\eqref{summary:pdf}. However this scaling form is not expected to be valid for larger fluctuations of $y_{\max}$ of $O(1)$ around its mean. For this one needs to study the atypical fluctuations which is done in the Section.~\ref{dist_aty}. Below we first provide some understanding of the exponent $\eta_k$ using a Hessian theory and a scaling argument.  

%\subsubsection{Understanding the exponent $\eta_k$}
%\label{hessianeta}
%~\\
\vspace*{0.5cm}
\noindent{\it Estimating the exponent $\eta_k$}
\vspace*{0.5cm}

\noindent
\textit{Hessian Approach:} Since the inverse temperature $\beta \sim O(1)$, one would expect that the small fluctuations of $y_{\max}$ around its mean can be described by making a quadratic approximation of the Hamiltonian characterised by a Hessian evaluated around the minimum energy position configuration $y^*_i = x^*_i/L_N$ for $i=1, 2, .. N$. This configuration $\{x^*_i\}$ can be obtained by minimizing the energy function in Eq.~\eqref{microhami} numerically using the Broyden-Fletcher-Goldfarb-Shanno (BFGS) algorithm~\cite{Fletcher_computer_1970,BROYDEN_IMA_1970}. Under the Hessian approximation the Hamiltonian takes the form
\begin{equation}\label{hess:hami}
E_k(\{x_i\}) \approx E_k(\{x^*_i\}) + \frac{1}{2}\sum_{i, j =1}^{N} H_{ij} (x_i - x^*_i)(x_j - x^*_j),
\end{equation}
where the Hessian matrix is given by
\begin{equation}\label{hess0}
	\begin{split}
	H_{ij}&=\Big [  \frac{\partial^2{E_k(\{x_i\})}}{\partial{x_i}\partial{x_j}}\Big]_{\{x^*_i\}}=\delta_{ij} \Big [ 1+ \sum_{n \neq i}^{N} \frac{J \text{sgn}(k)k(k+1)}{ \big (x^*_i-x^*_n \big )^{k+2}}\Big ] -(1-\delta_{ij})\frac{J \text{sgn}(k) k(k+1)}{ \big (x^*_i-x^*_j \big )^{k+2}}.
	\end{split}
\end{equation} 
The above Hessian approximation is justified when the standard devaition ($\sigma_{\Delta_i}$ where $\Delta_i = x_{i+1}-x_i$) of the $i^{\rm th}$ bond is smaller than the mean length of the bond ($\langle \Delta_i \rangle$) i.e. the relative fluctuations are very small ($\sigma_{\Delta_i}/\Delta_i \ll 1$) at a given temperature and system size $N$. We can invetigate the properties of interest using the Hessian Hamiltonian [Eq.~\eqref{hess:hami}] as a starting point. In other words one can perform MC simulations for the Hessian Hamiltonian [Eq.~\eqref{hess:hami}], which in principle allows for crossing. However, in the temperature regime considered here such events are very rare. Thus, assuming that the particles stays ordered, the variance of $y_{\max}=x_N/L_N$ is given by
\begin{equation}\label{hess1}
\sigma^2_{y_{\max}} = \frac{[H^{-1}]_{NN}}{L^2_N},
\end{equation}
where $H^{-1}$ is the inverse of the matrix $H$. We numerically perform this inversion and find that $\sigma_{y_{\max}}$ has the following $N$ scaling: 
\begin{equation}\label{hess2}
\sigma_{y_{\max}} \sim N^{-\eta_k^{\rm (hess)}}.
\end{equation}
We compare this exponent $\eta_k^{\rm (hess)}$ (obtained by inversion of the Hessian matrix) with the exponent obtained using MC simulations of both the original confined Riesz gas [Eq.~\eqref{microhami}] denoted by  $\eta_k$ [Eq.~\eqref{eta-k}] and the Hessian Hamiltonian [Eq.~\eqref{hess:hami}] denoted by $\eta_k^{\rm (hMC)}$ in Fig.~\ref{fig:etakvsk}. We observe an excellent agreement suggesting $\eta_k = \eta_k^{\rm (hess)}=\eta_k^{\rm (hMC)}$. The fact that $\eta_k^{\rm (hess)}=\eta_k^{\rm (hMC)}$ justifies the above assumption of almost non-crossing trajectories of particles at $O(1)$ temperature. While the Hessian theory along with the assumption of ordering implies the Gaussian form for the scaling distribution $\mathcal{F}_{\beta}^{(k)}(z)$, the actual MC simulation gives a non-Gaussian form as shown in Fig.~\ref{fig:distr}, even though the scale $\sigma_{y_{\max}}\sim N^{-\eta^{\rm (hess)}_k}$ of the data collapse is provided by the Hessian theory. Our findings therefore indicate that the Hessian theory (albeit an approximation) encodes some non-trivial features of the underlying confined Riesz gas. \\

\noindent
\textit{Analytical estimate of $\eta_k$ for $-2<k<0$:}
In order to find analytical estimate of $\eta_k$, we look at the relevant length scales present in the system. Two length scales $\ell_l$ and $\ell_r$ can be identified by estimating the distance (measured from $w = l_k^{\rm uc}$ on the left and right, respectively) at which the PDF starts having the exponential form while still being of $O(N^0)$. At these length scales the large deviation behaviour characterised by LDF $\Phi_{-}(w, k)$ and $\Phi_{+}(w, k)$ start becoming valid.  To identify $\ell_l$ we rewrite the probability in the left large deviation part in Eq.~\eqref{summary:pdf} as
\begin{equation}\label{ell_l}
\mathcal{P}_k\left[y_{\max}=w, N\right] \approx
        \exp\Bigg[{-\beta \left(\frac{l_k^{\rm uc}- w}{\ell_l}\right)^{e_k^{-}} }\Bigg],~\text{for}~ w \to l_k^{\rm uc-} ~\text{where}~ \ell_l \sim N^{-\frac{2 \alpha_k+1}{e_k^{-}}}.
\end{equation}
In Eq.~\eqref{ell_l}, we have used the asymptotic form of $\Phi_-(w, k)$ from Eq.~\eqref{philasym}. Note that when we approach $l_k^{\rm uc}$ from $w \ll l_k^{\rm uc}$, $\ell_l$ is the length scale at which the PDF in the left large deviation regime becomes $O(1)$ i.e $\mathcal{P}_k\left[w=l_k^{\rm uc}-\ell_l, N\right] \sim e^{-\beta N^{2 \alpha_k+1} \Phi_-(w, k) } \sim O(1)$.  Using a similar argument one can estimate
\begin{equation}\label{ell_r}
\ell_r \sim N^{-\frac{2 \alpha_k}{e_k^+}}.
\end{equation}
For $-2<k<-1$ and $k>0$ we find that $\ell_r < \ell_l$ while for $-1<k<0$ we find that $\ell_l<\ell_r$. Notice that at the smallest length scale, the PDF described by the left and right LDF is $O(N^0)$. %We further assume that at this length scale the large deviation behaviour smootly connects to the tails of the distribution of the typical fluctuations. Under this assumption, we arrive at the following conjecture of the exponent $\eta_k$ (characterizing the variance of $y_{\max}$)}
If this length scale is required to describe the typical fluctuation of $y_{\max}$, the LDF form of the PDF should smoothly match the tails of the distribution in the typical regime. Assuming that such smooth matching occurs at this scale, we arrive at the conjecture of the exponent
\begin{equation}\label{conjecture}
\eta_k^{\rm (c)} =
\begin{cases}
\frac{2 \alpha_k}{e_k^+} ~~~= \frac{4}{(k+2)(3-k)} & ~\text{ for }~-2<k<-1\\
\frac{2 \alpha_k+1}{e_k^-} = \frac{4+k}{3(k+2)} & ~~\text{for}~~~-1<k<0.
\end{cases}
\end{equation} 
where we have used the values of $e_k^+$ and $e_k^-$ which are calculated in Section.~\ref{distofxmax} and summarized in Table~\ref{table:epmk}.
This conjecture in Eq.~\eqref{conjecture} (solid line) agrees remarkably well with our numerical data as shown in Fig.~\ref{fig:etakvsk}. This excellent agreement (for $k<0$) verifies the presence of a single scale that smoothly connects the large devaition and the typical fluctuation regimes. The existence of a single scale is consistent with the fact that the field theory for $k<0$ is exact (in the sense that there are no subleading corrections in $N$) and moreover there is just a single term of $O(N^2)$ (the double sum in Eq.~\eqref{microhami} can be replaced by a double integral without invoking the notion of principal value).

For $k>0$, the argument based on the existence of a single scale connecting the large deviation and the typical regimes gives $\eta_k^{\rm (c)} =4/(k+2)(3-k)$ for $0<k<1$ and $\eta_k^{\rm (c)} =2 k/(k+2)$ for $k>1$.
%\begin{equation}\label{conjecturekg0}
%\eta_k^{\rm (c)} =
%\begin{cases}
%\frac{4}{(k+2)(3-k)} & ~\text{ for }~0<k<1\\
%\frac{2 k}{(k+2)} & ~~\text{for}~~~k>1.
%\end{cases}
%\end{equation}
We find that these values of the exponent for $k>0$ fail to describe the typical fluctuations. %\sout{We speculate the reason being the following: Since the actual scale (obatined from MC simulation) of fluctuation [$\sim O(N^{-\eta_k})$] is smaller than the mean gap [$\sim O(N^{-a_k^{\rm (e)}})$] [see Fig.~\ref{fig:etakvsk}], the leading order field theory (which in principle should correctly describe the fluctuations of the mean gap) is insufficient to capture the correct scaling}. 
This is probably due to the fact that the field theory for $k>0$ has the subleading corrections (higher order derivatives in density) in $N$, including the correction due to entropy. This could lead to multiple intermediate scales. %\sout{In order to capture the behaviour at smaller scale one requires to consider the subleading corrections to the free energy (higher order derivatives), including the correction due to entropy. This suggests that the system has multiple intermediate scales.}
 %As a consequence, the argument [Eq.~\eqref{conjecturekg0}] based on identifying shorter length scales from our large-$N$ field theory (which does not contain subleading terms) fails for $k>0$.

\subsection{Distribution in the atypical region}
\label{dist_aty}
As mentioned earlier, performing the multiple integrals in Eq.~\eqref{Z_N(W)} in terms of the microscopic variables $x_i$'s is a daunting task. Nonetheless, one can make progress by converting it to a problem of functional integration over densities and then employing the saddle point calculation in the large $N$ limit. For a given configuration of the positions $(x_1,x_2,...,x_N)$ of the particles, we define the empirical density as 
\begin{equation}
    \hat{\rho}_N(x) = \frac{1}{N} \sum_{i=1}^N \delta\left(x-x_i\right) \;. \label{rho_emp}
\end{equation}
We perform the multiple integrals in two steps.
\begin{enumerate}
\item Integrate over the microscopic configurations corresponding to a given macroscopic density profile $\rho_N(x)$. This introduces an entropy contribution corresponding to this density profile. Also, this stage involves converting the energy function $E_k(\{x_i\})$ in Eq.~\eqref{microhami} to an energy functional $\mathcal{E}_k[\rho_N(x)]$~\cite{agarwal2019harmonically,kumar2020particles}. This yields
\begin{equation}
  Z_k(W, \beta, N) \approx \int D[\rho_N] e^{-\beta \mathcal{E}_k\left[\rho_N(x)\right]-  N\int dx \,\, \rho_N(x)\ln\left(\rho_N(x)\right) } \delta \left(\int_{-\infty}^{W} dx \rho_N(x)-1 \right),     \label{Z_k-func}
\end{equation}
where the delta function ensures that the functional integrals are performed only over normalised density profiles. For large $N$, it has been shown~\cite{agarwal2019harmonically, kethepalli2021harmonically} that the energy functional $\mathcal{E}_k[\rho_N]$ takes the following form, depending on the value of $k$
\begin{equation}\label{mcal(E)}
    \mathcal{E}_k\left[\rho_N(x)\right] \approx \frac{N}{2} \int_{-\infty}^W dx \,\, x^2 \rho_N(x) + 
    \begin{cases}
     \zeta(k) N^{k+1}\int_{-\infty}^W dx \,\, \left[\rho_N(x)\right]^{k+1},& \text{for}\quad k>1\\
      & \\
      N^{2}\ln N\int_{-\infty}^W dx \,\, \left[\rho_N(x)\right]^{2},& \text{for}\quad k=1\\
      & \\
     \frac{\text{sgn}(k)N^2}{2}\int_{-\infty}^W dx' dx \frac{\rho_N(x')\rho_N(x)}{|x-x'|^k} 
      &\text{for}-2<k<1.
    \end{cases}
\end{equation}
\item We perform the functional integral in Eq.~\eqref{Z_k-func} using the saddle point method. We further find the saddle point density by extremizing the action which finally provides $Z_k(W, \beta, N)$ in the exponential form, as will be shown later. 
\end{enumerate}
To proceed, it is convenient to use the scaled position variables $y_i = x_i/L_N$ where $L_N$ is the length scale over which the density profile $\rho_N(x)$ varies.
To establish the $N$ dependence  of $L_N$ given in Eq.~\eqref{scaling}, we rescale the density as $\rho_N(x)=L_N^{-1}{\rho}_k\left({x}{  L_N^{-1}}\right)$ and substitute this scaling form
in Eq.~\eqref{mcal(E)}. The first term corresponding to the confining harmonic potential scales as $N \, L_N^2$. The scaling of the interaction term depends on $k$. For $k >1$, it scales as $N^{k+1}\, L_N^{-k}$, while for $k=1$ it scales as  $N^2 \, (\ln N)/L_N$ and for $-2<k<1$ it scales as $N^2 L_N^{-k}$\;. Matching the interaction term and the confining term (for a given $k$), one finds that $L_N \sim N^{\alpha_k}$ where $\alpha_k$ is given in Eq.~\eqref{scaling}. Following a similar computation one can show that $L_N = (N \, \ln N)^{1/3}$ for the marginal case $k=1$. Plugging the scaling form for $\rho_N(x)$ in Eq. (\ref{mcal(E)}), one gets~\cite{agarwal2019harmonically}
\begin{align}
 \mathcal{E}_k\left[{\rho}_N(x)\right] = \mathcal{B}_N
\tilde{\mathcal{E}}_k\left[{\rho}_k(x L_N^{-1})\right],~\text{where} ~~
\mathcal{B}_N=
\begin{cases}
N^{2 \alpha_k +1}~&\text{for}~k \neq 1 \\ 
N^{5/3} (\ln N)^{2/3}~&\text{for}~k = 1,
\end{cases}
\label{scale-mcal(E)}
 \end{align}
 and the scaled energy functional $ \tilde{\mathcal{E}}_k\left[{\rho}_k(y)\right]$ takes the following forms
\begin{equation}\label{scaled-field}
    \tilde{\mathcal{E}}_k\left[\rho_k(y)\right] \approx \frac{1}{2} \int_{-\infty}^w dy \,\, y^2 \rho_k(y) +
    \begin{cases}
  \zeta(k) \int_{-\infty}^w dy \,\, \rho_k(y)^{k+1}& \quad \quad \;\; k> 1\\
   \int_{-\infty}^w dy \,\, \rho_k(y)^{2}& \quad \quad \;\; k=1 \\
 \frac{ \text{sgn}(k)}{2} \int_{-\infty}^w \int_{-\infty}^w dy' dy \,\, \frac{\rho_k(y)\rho_k(y')}{|y'-y|^k} & -2<k<1 \,,
    \end{cases}
\end{equation}
with $w=W/L_N$. Substituting (\ref{scale-mcal(E)}) in the expression (\ref{Z_k-func}) for the partition function, one finds that for $k>-2$ the energy scale
$\mathcal{B}_N$ is much bigger than the scale of the entropy, since $\mathcal{B}_N \gg N$ for large $N$ and $\beta L_N^2 > 1$. At high temperatures $T \sim O(N^{2\alpha_k})$, the energy and entropy terms become comparable and the free energy gets modified. Neglecting the entropy term and using the integral representation of the delta function $\delta(x) = \int_\Gamma \frac{d\mu}{2\pi i}e^{\mu x}$ where $\Gamma$ runs along the imaginary axis in the complex $\mu$-plane, we rewrite the  partition function in Eq.~\eqref{Z_k-func} as 
\begin{equation}\label{partition-field}
    Z_k(w L_N, N) = \int d\mu \int \mathcal{D}[\rho_k]~\text{exp}\left[-\beta  \mathcal{B}_N \Sigma_k\left[ \rho_k(y),\mu\right] + o(\mathcal{B}_N)\right] ,
\end{equation}
where the action $\Sigma_k\left[\rho_k(y),\mu\right]$ is given by 
\begin{equation}\label{action}
    \Sigma_k\left[\rho_k(y),\mu\right] = \left(\tilde{\mathcal{E}}_k\left[\rho_k(y)\right] - \mu \left(\int dy \rho_k(y) -1\right)\right) \;,
\end{equation}
with $\tilde{\mathcal{E}}_k\left[\rho_k(y)\right]$ given in Eq. (\ref{scaled-field}).  

The integrals in Eq.~\eqref{partition-field} can be performed using the saddle point method in which one requires to minimise the action $\Sigma_k[{{\rho}}_k(y), \mu]$ in Eq. (\ref{action}) to find the saddle point density $\rho^*_k(y,w)$ and the chemical potential $\mu_k^*(w)$. The saddle point equations then read
\begin{align}
    \frac{\delta \Sigma_k\left[\rho_k\left(y\right), \mu \right]}{\delta \rho_k\left(y\right)}\Bigg|_{\substack{\rho_k\left(y\right) = \rho^*_k\left(y, w\right)\\\mu=\mu_k^*(w)}} &= 0, \label{saddle-den_eqn}\\
    \frac{\partial \Sigma_k\left[\rho_k\left(y\right), \mu \right]}{\partial \mu}\Bigg|_{\substack{\rho_k\left(y\right) = \rho^*_k\left(y, w\right)\\\mu=\mu_k^*(w)}} &= 0 \;. \label{saddle-mu_eqn}
\end{align}
Note that the second equation above is precisely the normalization condition $\int dy \rho_k(y) = 1$. 
Solving the above two equations \eqref{saddle-den_eqn} and \eqref{saddle-mu_eqn} satisfying the normalisation condition, one finds $\rho^*_k(y,w)$. In the limit $N \to \infty$, the saddle point density $\rho^*_k(y,w)$ is actually   the average density of the particles of the Riesz gas in the presence of a hard wall at $W=wL_N$. In a recent work, explicit expressions for the saddle point density $\rho^*_k(y,w)$ (here after called as constrained densities) have been obtained for all values of $k>-2$ \cite{kethepalli2021harmonically}.

Substituting this saddle point density  in Eqs.~\eqref{partition-field} and \eqref{action}, one finds the partition function $Z_k(W,\beta,N)$ in Eq.~\eqref{Z_N(W)} as $Z_k(W,\beta,N) \approx \exp \left[ -\beta \mathcal{B}_N \tilde{\mathcal{E}}_k\left[\rho^*_k(y, w)\right] \right]$ where the (scaled) energy functional is given in Eq.~\eqref{scaled-field}. Hence, using Eq.~\eqref{CDF:ratio}, the CDF of the position of the rightmost particle is given by 
\begin{equation}\label{PDF:leftld}
 {\rm Prob.}[x_{\max}<w L_N, N] \approx \exp{\left[-\beta \mathcal{B}_N \left(\tilde{\mathcal{E}}_k\left[\rho^*_k(y, w)\right]-\tilde{\mathcal{E}}_k\left[\rho^*_k(y, w\to \infty)\right] \right)\right]},
\end{equation}
where $w=W/L_N$ represents the scaled position of the wall. Notice that in the limit $w \to \infty$, the saddle point density $\rho^*_k(y, w)$ corresponds to the density of the unconstrained gas i.e. in the absence of any wall. As mentioned earlier, this unconstrained density $\rho_{k, {\rm uc}}^{\rm{*}}(y) = \rho^*_k(y, w\to \infty)$ was computed in Ref.~\cite{agarwal2019harmonically} where it was shown to be given by Eq.~\eqref{rho_uc}.

Note that  if the wall is placed outside the support  of the unconstrained density (i.e. $w>l_k^{\rm uc}$) the density profile remains unchanged, because $\rho_{k, {\rm uc}}^{\rm{*}}(y)$ has a finite support $[-l_k^{\rm uc}, l_k^{\rm uc}]$. In other words, $ \rho^*_k(y, w) = \rho_{k, {\rm uc}}^{\rm{*}}(y)$ for $w\geq l_k^{\rm uc}$  since the effect of the hard wall is noticeable only when $w<l_k^{\rm uc}$. Consequently, the right hand side of Eq.~\eqref{PDF:leftld} can not describe the probability distribution of $y_{\max}$ for $y_{\max} >  l_k^{\rm uc}$. In this case one needs to employ a different method to find the CDF of $y_{\max}$. This is expected since, intuitively, one would anticipate to have different energy costs for creating a fluctuation with $y_{\max} < l_k^{\rm uc}$ and $y_{\max} > l_k^{\rm uc}$.

To compute the PDF for $y_{\max} > l_k^{\rm uc}$, we follow the procedure described in Ref.~\cite{majumdar2009large}. In Ref.~\cite{majumdar2009large} it was argued for the Dyson's log-gas that for large $N$ the dominant contribution to the PDF for $y_{\max} > l_k^{\rm uc}$ would come from the energy cost required to pull the rightmost particle to the right of the right edge of the unconstrained density. Assuming the same mechanism to hold for all values of $k>-2$ (which will be verified numerically later), we write
\begin{equation}\label{PDF:rightld}
\begin{split}
{\rm Prob.}[x_{\max}=W, N] 
%&= {\rm Prob.}[x_N=W, \{x_i<W\}_1^{N-1}, \beta, N]\\&
&=  \frac{1}{Z_k(\beta,N-1)}\int_{-\infty}^{W} dx_{N-1}  \ldots \int_{-\infty}^{x_2} dx_{1} \exp(-\beta E(x_1,x_2,...,x_{N-1}))\\&
\quad \quad \quad \times \exp\left[-\beta\left(\frac{W^2}{2}+ {\rm sgn}(k) \sum_{i=1}^{N-1} |W-x_i|^{-k}\right)\right]\\&
%=\left\langle exp\left[-\beta\left(\frac{W^2}{2}+ {\rm sgn}(k) \sum_{i=1}^{N-1} |W-x_i|^{-k}\right)\right]\right\rangle_{N-1}\\&
=\left\langle \exp\left[-\beta\left(\frac{W^2}{2}+ {\rm sgn}(k) N \int dx \frac{\rho_{N-1}(x)}{|W-x|^{k}} \right)\right]\right\rangle_{N-1},
\end{split}
\end{equation}
where $E(x_1,x_2,...,x_{N-1})$ represents the energy of the $(N-1)$ particles in the absence of the rightmost particle and the angular brackets $\langle . \rangle_{N-1}$ denotes the average with respect to $N-1$ particle distribution. To compute this average we again use the saddle point method. For large-$N$, the difference between the densities of $N-1$ particles and $N$ particles is negligible. Hence the saddle point density is given by the unconstrained density  in Eq.~\eqref{rho_uc}. Using this  in Eq.~\eqref{PDF:rightld} and expressing the PDF in terms of the scaled variable $w = W/L_N$, we get 
\begin{equation}\label{PDF:right}
\begin{split}
{\rm Prob.}[x_{\max}=w L_N, N] &\approx \exp \Bigg[-\beta L_N^2 \Bigg(\frac{w^2-{l_k^{\rm uc}}^2}{2} \\ 
& \left. \right. +\frac{ {\rm sgn}(k) N}{L_N^{k+2}} \int dy\rho_{k,{\rm uc}}^{\rm *}(y) \left( \frac{1}{|w-y|^{k}} - \frac{1}{|l_k^{\rm uc}-y|^{k}}\right) \Bigg) \Bigg].
\end{split}
\end{equation}
%\quad \quad -exp\Bigg[\beta L_N^2 \Bigg(\frac{(l_k^{\rm uc})^2}{2}+\frac{ sgn(k) N}{L_N^{k+2}} \int dy \frac{\rho_{k,{\rm uc}}^{\rm *}(y)}{|l_k^{\rm uc}-y|^{k}} \Bigg)\Bigg]
The expressions in Eqs.~\eqref{PDF:leftld} and \eqref{PDF:right} suggest the following large deviation form of the CDF of $y_{\max} = x_{\max}/L_N$
\begin{align}
 {\rm Prob.}[y_{\max}<w, N] \approx 
 \begin{cases}
 \exp(-\beta \mathcal{B}_N \Phi_-(w,k)), & {\rm for}~l_k^{\rm uc} - w \geq O(1) \\
 1-\exp(-\beta L_N^2 \Phi_+(w,k)), & {\rm for}~w-l_k^{\rm uc}  \geq O(1),
 \end{cases}
\end{align}
in the large $N$ limit, where $\mathcal{B}_N$ and $L_N$ are given in Eqs.~\eqref{scale-mcal(E)} and \eqref{scaling}, respectively.
The LDF are  given by
\begin{equation}
\begin{split}
\Phi_-(w,k) &= \tilde{\mathcal{E}}_k\left[\rho^*_k(y, w)\right]-\tilde{\mathcal{E}}_k\left[\rho_{k,{\rm uc}}^{\rm *}(y)\right], \\
 \Phi_+(w, k) &= \left(\frac{w^2-{l_k^{\rm uc}}^2}{2}\right) + {\rm sgn}(k)\Theta(1-k) \int dy\rho_{k,{\rm uc}}^{\rm *}(y) \left( \frac{1}{|w-y|^{k}} - \frac{1}{|l_k^{\rm uc}-y|^{k}}\right). 
\end{split}
\label{LDF-general}
\end{equation}
The form of the saddle point densities $\rho^*_k(y, w)$ and $\rho_{k,{\rm uc}}^{\rm *}(y)$, in the presence and the absence of the wall, respectively, depends explicitly on $k$. Therefore the form of the LDF as well as the exponents $e^{\mp}_k$ characterising their asymptotic behaviours near the wall, would also depend on $k$. In the following, we compute the explicit form of the LDF and the exponents in the following regimes (1) $k\geq1$ (2) $-1<k<1$ and (3) $-2<k<-1$, separately.

\subsubsection{Regime 1: Short-ranged interactions  $(k \geq 1)$ -}
\label{sec:regime1}
\begin{figure}[h!]
    \centering
    %\quad \quad \quad ~~\includegraphics[scale=0.9]{img/wdens_1.png}
    \quad \quad \quad ~~\includegraphics[scale=0.6]{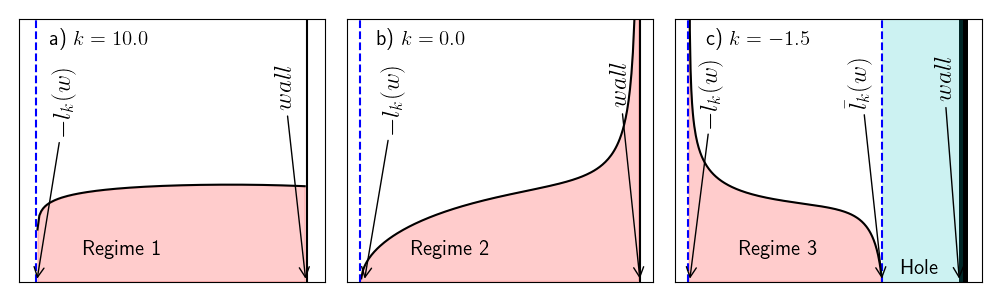}
    \caption{A plot of the scaled average density in the presence of wall $\rho_{k}^{\rm{*}}(y, w)$ versus $y$ for the three regimes (a) $k\geq 1$, (b) $-1< k < 1$ and (c) $-2 < k \leq -1$. The blue dashed vertical line indicates the left edge [$-l_k(w)$] of the support and the black solid line represents the wall position $w$. In regime $1$ the density is constant at the wall while it diverges in regime $2$ . In both of these regimes the density vanishes at the left edge. In the regime 3 the density has two disjoint regions, an extended bulk part [$-l_k(w)<y<\bar{l}_k(w)$] and a delta function at the wall position (shown by a thick solid vertical line). They are separated by a hole region [$\bar{l}_k(w)<y<w$] devoid of particles (shaded cyan region).}
    \label{fig:wdens}
\end{figure}
%(for $k=10$ in (a), the true density at the left edge vanishes, though it is not clearly visible due to the compression of the scale)

\begin{figure}[h!]
    \centering
    \includegraphics[scale=0.7]{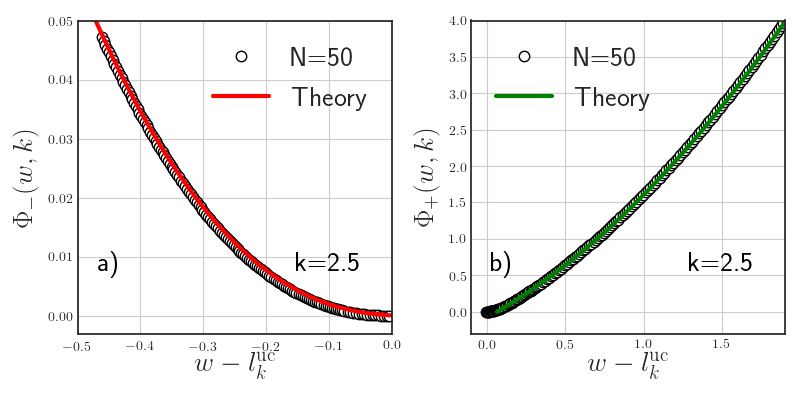}
    \caption{\textit{Regime 1 ($k \geq 1$):} The numerical verification of the LDF $\Phi_{\pm}(w, k)$ given in Eq.~\eqref{ldf:lkg1_1} and Eq.~\eqref{ldf:rkg1_1}, respectively in panels (a) and (b). The rare events such that $|y_{\max}-l_k^{\rm uc}| \sim O(1)$ are generated using the importance sampling method~\cite{hartmann2011large, hartmann2018distribution} and the associated probabilities are computed from which the large deviation functions are calculated numerically. The parameters used in the simulations are $J=1$ and $\beta=1$.}
    \label{fig:k2p5}
\end{figure}
In this regime the interaction energy falls relatively fast with increasing separation, i.e. it  effectively acts as short-ranged. Consequently, the energy functional, given in Eq.~\eqref{scaled-field},
is local in the leading order for large $N$. Using this functional one finds that the saddle point equation Eq.~\eqref{saddle-den_eqn} becomes
\begin{equation}\label{kg1-chem-eq}
 \mu_k^*(w) = \frac{y^2}{2}  + (k+1) \zeta(k)\left[\rho^*_k \left(y, w \right) \right]^{k},
\end{equation}
which one needs to solve with the normalization condition obtained from Eq.~\eqref{saddle-mu_eqn}.  
The density is then given by~\cite{kethepalli2021harmonically} [see Fig.~\ref{fig:wdens}a]
\begin{equation}\label{kg1densii}
    \rho^*_k(y, w) = A_k \left(l_k(w)^2 - y^2\right)^{\frac{1}{k}}  \;,\; -l_k(w) \leq y \leq w \;. 
\end{equation}
Substituting this form of the density profile in the normalization condition gives us an equation for $l_k(w)$ expressed in terms of an auxiliary variable 
\begin{equation}\label{mkw}
m_k(w)= \frac{w+l_k(w)}{2 l_{k}(w)},~\text{for}~k \geq 1 ~\text{and}~w\leq l_k^{\rm uc},
\end{equation}
as
\begin{align}
(2m_k(w)-1) \left(\frac{B(\gamma_k +1, \gamma_k+1)}{B(m_k(w); \gamma_k +1, \gamma_k+1)}\right)^{\alpha_k}=\frac{w}{l_k^{\rm uc}}.
  \label{l_{k}(w)-k>1}
\end{align}
%where $B(m_k(w); a, b) = \int_0^{m_k(w)} u^{a-1}(1-u)^{b-1}\, du$ is the incomplete Beta function and 
We recall that in this regime $\gamma_k = 1/k$ and $\alpha_k = k/(k+2)$. The variable $m_k(w)$ lies in the range $[0,1]$. Solving Eq. (\ref{l_{k}(w)-k>1}) gives $m_k(w)$, which in turn fixes the left edge of the support $l_k(w)$ through Eq.~\eqref{mkw}. Note that in this regime the density at the wall is finite.

Finally by substituting the density profile $\rho^*_k(y, w)$ in Eq.~\eqref{scaled-field} we get the scaled energy functional $\tilde{\mathcal{E}}_k[\rho^*_k]$, which together with Eq.~\eqref{LDF-general} allows to compute the LDF $\Phi_{\pm}(w, k)$.  We find the following explicit expressions (see~\ref{apndkg1} for details)
\begin{equation}\label{ldf:lkg1_1}
\begin{split}
\Phi_{-}(w, k) &= \frac{l_k(w)^2}{2(k+1)} \Bigg[1+ \frac{4 k}{B(m_k(w),\gamma_k+1,\gamma_k+1)} \bigg(B(m_k(w), \gamma_k+3, \gamma_k+1)\\&-B(m_k(w), \gamma_k+2, \gamma_k+1)+ \frac{B(m_k(w), \gamma_k+1, \gamma_k+1)}{4}\bigg)\Bigg]-\frac{(l_k^{\rm uc})^2(k+2)}{2(3k+2)},
\end{split}
\end{equation}
\begin{equation}
\Phi_{+}(w, k) = \frac{w^2 - \left(l_k^{\rm uc}\right)^2}{2}\label{ldf:rkg1_1}.
\end{equation}
Here $B(x,a,b)$ is the incomplete Beta function.

Using an importance sampling method described in the~\ref{apndims}, we compute the probability distribution $P_{\rm num}(y_{\max})$ which includes the atypical part also. To extract the left large deviation function we plot $-\log{\big(P_{\rm num}(y_{\max})\big)}/\mathcal{B}_N$ as a function of $y_{\max}-\langle y_{\max} \rangle$. Similarly the right large deviation function is extracted by plotting $-\log{\big(P_{\rm num}(y_{\max})\big)}/L_N^2$ as a function of $y_{\max}-\langle y_{\max} \rangle$. In Fig.~\ref{fig:k2p5}, we compare the LDF obtained numerically with our analytical expression given in Eqs.~\eqref{ldf:lkg1_1} and~\eqref{ldf:rkg1_1} and observe remarkable agreement up to an overall translation on the $x$ axis. This translation is an artifact of finite size effect, due to which $\langle y_{\max} \rangle$ is slightly different from its theoretical value $l_k^{\rm uc}$ in the thermodynamic limit.

We study the asymptotic behaviour of $\Phi_{-}(w, k)$ as $w\to l_k^{\rm uc-}$. From Eq.~\eqref{l_{k}(w)-k>1} we obtain the asymptotic behavious of $m_k(w)$ as $w \to l_k^{\rm uc-}$ as 
\begin{equation}\label{kg1:epsi}
\begin{split}
 m_k(w) &\approx 1-\frac{l_k^{\rm uc}-w}{2l_k^{\rm uc}} + o(l_k^{\rm uc}-w). %O((l_k^{\rm uc}-w)^2).
\end{split}
\end{equation}
Performing the series expansion about $m_k(w) = 1$ and using the approximation of $m_k(w)$ [Eq.~\eqref{kg1:epsi}] we get from Eq.~\eqref{ldf:lkg1_1}
\begin{equation}\label{phimekmkg1}
 \Phi_-(w, k) \approx \frac{k^2(2l_k^{\rm uc})^{-\frac{1}{k}} B(1+\gamma_k,1+\gamma_k)}{2(2k+1)(k+1)} \left(l_k^{\rm uc}-w \right)^{e^-_k} \text{   with   } e^-_k = 2+\frac{1}{k}.
\end{equation}
Hence the system undergoes a $3^{\rm rd}$ order phase transition based on Ehrenfest classification. The asymptotic behaviour of $\Phi_{+}(w, k)$ can be obtained by taking the limit $w\to l_k^{\rm uc+}$ in Eq.~\eqref{ldf:rkg1_1} and is given by
\begin{equation}\label{phipekpkg1}
 \Phi_+(w, k) \approx l_k^{\rm uc} \left(w-l_k^{\rm uc} \right)^{e^+_k} \text{   with   } e^+_k = 1.
\end{equation}

\subsubsection{Regime 2: Weakly long-ranged interactions $(-1 < k < 1)$ -}
\label{sec:regime2}
\begin{figure}[h!]
    \centering
    \includegraphics[scale=0.7]{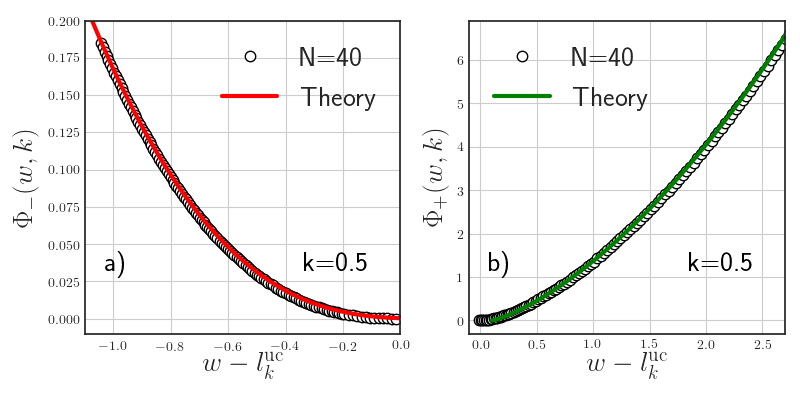}
    \caption{\textit{Regime 2 ($-1<k < 1$):} The numerical verification of the LDF $\Phi_{\pm}(w, k)$ given in Eq.~\eqref{ldf:lkgm1_2} and Eq.~\eqref{ldf:rkgm1_3}, respectively in panels (a) and (b). The probabilities of rare events such that $|y_{\max}-l_k^{\rm uc}| \sim O(1)$ are computed numerically from which the large deviation functions are extracted. The parameters used in the simulations are $J=1$ and $\beta=1$.}
    \label{fig:k0p5}
\end{figure}

In this regime of $k$, the interaction forces decay slower with increasing inter-particle separation compared to the previous short-ranged regime. The energy functional in this regime is given in Eq.~\eqref{scaled-field} and  is non-local in leading order for large $N$. Using this energy functional in Eq.~\eqref{saddle-den_eqn}, we obtain the saddle point equation as 
\begin{equation}\label{kl1 chem eq}
\mu_k^*(w) = \frac{y^2}{2}  + \text{sgn}(k)\int_{-\infty}^{w} dy' \,\, \frac{\rho^*_k\left(y', w\right)}{|y'-y|^k}.
\end{equation}
This Eq.~\eqref{kl1 chem eq} has been solved recently~\cite{kethepalli2021harmonically} using the Sonin inversion formula~\cite{buldyrev2001properties} and is given by [see Fig.~\ref{fig:wdens}b]
\begin{equation}\label{kl1dens}
{\rho}^* _k(y, w) =A_k (l_{k}(w)+y)^{\frac{k+1}{2}} (w-y)^{\frac{k-1}{2}} \left(\tilde{l}_k(w)-y \right), ~~ \text{for} ~~ -l_{k}(w) \leq y \leq w, %\;,\; ~~ w<l_k^{\rm uc},
\end{equation}
where $\tilde{l}_k(w) = \frac{1}{2}\big((k+1) l_{k}(w)+(1-k) w\big)$. Here, it is worth noting that the density at the wall has an integrable divergence while it vanishes on the left edge $-l_k(w)$ of the support. The quantity $l_k(w)$ is determined from the normalization condition $\int_{-l_{k}(w)}^w \rho^*_k(y, w)\, dy =1$ which leads to~\cite{kethepalli2021harmonically}
\begin{align}
 \label{l_{k}(w)-k<1}
 \left(\frac{k+3-2g_k(w)}{k+1}\right)\left(\frac{2g_k(w)\left(2+k\right)-\left(k+3\right)}{k+1}\right)^{-\alpha_k} = \frac{w}{l_k^{\rm uc}},
\end{align}
where the auxiliary variable is
\begin{equation}\label{gkwkl1}
g_k(w)=\frac{l_k(w)+\tilde{l}_k(w)}{w+l_k(w)},~\text{  for  }~ -1 <k<1~\text{and}~w<l_k^{\rm uc}.
\end{equation}
We recall that $\alpha_k = 1/(k+2)$ [see Eq.~\eqref{scaling}]. This equation is the analogue of Eq.~\eqref{l_{k}(w)-k>1} in the regime 1. This equation gives $g_k(w)$ for a fixed $w$, which is used to find the left edge of the support $l_k(w)$ using Eq.~\eqref{gkwkl1}. We use this saddle point density to find the large deviation function given in Eq.~\eqref{LDF-general}. To do so we first need to calculate the scaled energy functional $\tilde{\mathcal{E}}_k[\rho^*_k]$ given in Eq.~\eqref{scaled-field}. We relegate some details of this computation in the~\ref{apndkl1}. Here we present only the final expressions namely
\begin{equation}
\begin{split}
\Phi_{-}(w, k) = & (k+2)(l_k^{\rm uc})^2\Bigg[ \left(1+\frac{2(k+2)}{k+1}(g_k(w)-1)\right)^{-\frac{k+4}{k+2}} \Big[\frac{1}{2k(k+4)}\\&+\frac{(g_k(w)-1)}{k(k+1)}+\frac{2(g_k(w)-1)^2}{k(k+1)^2} + \frac{4(g_k(w)-1)^3}{(1+k)^3}\Big] - \frac{1}{2k(k+4)}\Bigg]\label{ldf:lkgm1_2},
\end{split}
\end{equation}
\begin{align}
%\begin{split}
\Phi_{+}(w, k) = &\left(l_k^{\rm uc}\right)^2 \frac{32 (g_k^{\rm uc}(w)^{-1}-1)^{\frac{3-k}{2}}B(2+k, \frac{5-k}{2})}{(k+3)(k+5)(k+7)}\times \notag \\ & \quad \quad ~_2F_1 \big[-\frac{k+1}{2},\frac{k+3}{2}, \frac{5-k}{2}, 1-g_k^{\rm uc}(w)^{-1}\big],
\label{ldf:rkgm1_3}
%\end{split}
\end{align}
where $g_k^{\rm uc}(w) = 2 l_k^{\rm uc}/(w + l_k^{\rm uc})$ and
\begin{align}
~_2F_1[a,b,c,u] &= B(b, c-b)^{-1}\int_0^1 ds \frac{s^{b-1}(1-s)^{c-1-b}}{(1-us)^a} \\
&= \sum_{n=0}^{\infty} \frac{(a)_n(b)_n}{(c)_n} \frac{z^n}{n!}
%&= 1+ \frac{ab}{c} \frac{z}{1!}+ \frac{a(a+1)b(b+1)}{c(c+1)} \frac{z^2}{2!}\\& \notag + \frac{a(a+1)(a+2)b(b+1)(b+2)}{c(c+1)(c+2)} \frac{z^3}{3!}...
\end{align}
is the hypergeometric function with $(a)_n = a(a+1)(a+2)...(a+n)$ being the Pochhammer symbol. Note that $g_k^{\rm uc}(w)$ is the ratio of the size of the unconstrained gas to that of the constrained gas. These expressions of LDF are in excellent agreement with our numerical results obtained using the importance sampling method [see~\ref{apndims}] as can be seen in Fig.~\ref{fig:k0p5}. As mentioned earlier the LDF describe the pulled to pushed type phase transition and its nature is determined by the asymptotic behaviour of LDF near the right edge of the support of the unconstrained density. 

To characterize this asymptotic behaviour of LDF for $w \to l_k^{\rm uc-}$ we need to expand $g_k(w)$ around $w = l_k^{\rm uc}$. From the Eq.~\eqref{l_{k}(w)-k<1} we observe that
\begin{equation}\label{kl1:epsi}
\begin{split}
g_k(w) &\approx  1 + \frac{k+1}{4l_k^{\rm uc}}(l_k^{\rm uc}-w)  + o(l_k^{\rm uc}-w),\\
g^{\rm uc}_k(w) &=1+ \frac{w-l_k^{\rm uc}}{2 l_k^{\rm uc}}.
\end{split}
\end{equation}
We expand $\Phi_-(w, k)$ and $\Phi_+(w, k)$ in powers of $g_k(w)-1$ and $g^{\rm uc}_k(w)^{-1}-1$, respectively and then use these expansions Eq.~\eqref{kl1:epsi} which give
\begin{align}
\Phi_-(w, k) &\approx \frac{(k+2)}{12 l_k^{\rm uc}} \left(l_k^{\rm uc}-w \right)^{e^-_k}  \text{   with   } e^-_k = 3,\label{asym:ldf}\\
\Phi_+(w, k) &\approx \frac{2 (1-k) B\left(k+2,\frac{5-k}{2}\right)(l_k^{\rm uc})^{-e^+_k}}{3 A_k (k+5) (k+7) \left| k\right|  B\left(\frac{k+3}{2},1-k\right)}  \left(w- l_k^{\rm uc}\right)^{e^+_k}  \text{   with   } e^+_k = \frac{3-k}{2}\label{asym:rdf}.
\end{align}
The exponent $e_k^{-} = 3$ suggests that the system undergoes a $3^{\rm rd}$ order phase transition. 

\subsubsection{Regime 3: Strongly long-ranged interactions $(-2 < k \leq -1)$ -}
\label{sec:regime3}
\begin{figure}[h!]
    \centering
    \includegraphics[scale=0.7]{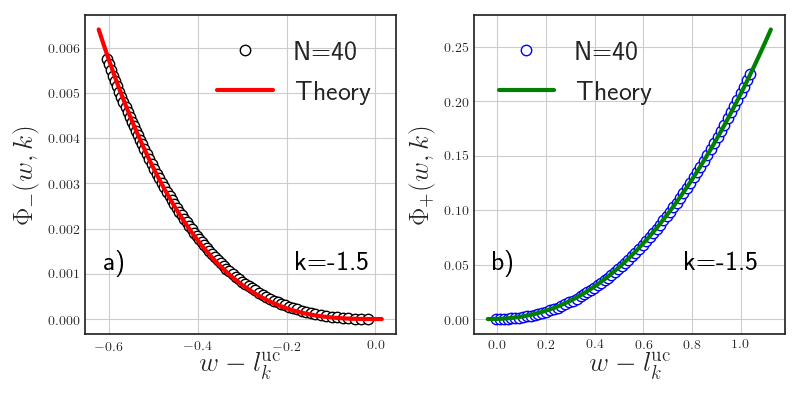}
    \caption{\textit{Regime 3 ($-2<k <-1$):} The numerical verification of the LDF $\Phi_{\pm}(w, k)$ given in Eq.~\eqref{ldf:lklm1_2} and Eq.~\eqref{ldf:rklm1_3}, respectively in panels (a) and (b). The probabilities of rare events such that $|y_{\max}-l_k^{\rm uc}| \sim O(1)$ are computed from which the large deviation functions are calculated numerically. In the simulation we use the parameters $J=1$ and $\beta=10^{-3}$. It is easy to show that our field theory calculation remains valid at this temperature since it satisfies the condition $\beta L_N^2>1$.}
    \label{fig:km1p5}
\end{figure} 

This regime is a bit more complicated since the constrained density has disjoint parts namely a delta function at the wall and an extended part separated by a region devoid of particles~\cite{kethepalli2021harmonically}. The divergence of the density at the wall seen in the previous regime (see Section.~\ref{sec:regime2}) becomes a delta function. This is rooted in the fact that the particles are allowed to sit at the same position. It turns out that due to the intricate interplay between the repulsive interaction and the confining harmonic potential the fraction of the particles tries to sit together at the wall. The rest of the particles then gets pushed away from the wall by ``super charge" resulting from the delta function. This creates a hole between the delta function and the extended region~\cite{kethepalli2021harmonically}. 
This density profile is obtained by solving the saddle point equations~\eqref{saddle-den_eqn}, which in this regime takes the form
\begin{equation}\label{klm1 chem eq}
\mu_k^*(w) = \frac{y^2}{2}  + \text{sgn}(k)\int_{-\infty}^{w} dy' \,\, \frac{\rho^*_k\left(y', w\right)}{|y'-y|^k}.
\end{equation}
This equation is solved in detail in Ref.~\cite{kethepalli2021harmonically} leading to the result [see Fig.~\ref{fig:wdens}c]
\begin{equation}\label{dens-s:klm1}
\rho^*_k(y,w) =     
    \begin{cases}
    \underbrace{A_k \frac{\left(l_{k}(w)+y\right)^{\frac{k+1}{2}} \left(\bar{l}_k(w)-y\right)^{\frac{k+3}{2}}}{(w-y)} 
     \mathbb{I}[-l_{k}(w) < y \leq \bar{l}_k(w)]}_{\rm extended}  &+ \, \underbrace{D^*_k(w)\delta(w-y)}_{\rm ``super~charge"}, \\
%     &\Theta(l_{k}(w)+y)\Theta(\tilde{l}_k(w)-y) \\ 
 ~~ & \quad \text{for} \quad w>w_c(k) \\
  %& \\
     ~~~~~~\delta(w-y), & \quad \text{for} \quad w<w_c(k),
    \end{cases}
\end{equation}
where $\mathbb{I}[a < z \leq b]$ represents the indicator function of the interval $[a, b]$. The amplitude $A_k$ is given in Eq.~\eqref{A_1k} and 
\begin{equation}
w_c(k) = \frac{(k+2) \left| k (k+1)\right| ^{\frac{1}{k+2}}}{k+1}.
\end{equation}
The other constants in Eq.~\eqref{dens-s:klm1} are expressed in term of the position $-l_k(w)$ of the left edge of the extended part of the density and are given by
\begin{align}
 %A_k&=\frac{\sin \left(\frac{\pi  (k+1)}{2} \right)}{(\pi  J (k+1) \left| k\right| )}, \\
 \bar{l}_k(w) &= \frac{2 w + (k+1) l_{k}(w)}{k+3}, \label{barl}\\
 D^*_k(w) &= \frac{\left(l_{k}(w)-w\right)\left(w+l_{k}(w)\right)^{\frac{k+1}{2}}}{ \left|k\right|\left(k+3\right)} \left(\frac{(k+1)\left(w-l_{k}(w)\right)}{k+3}\right)^{\frac{k+1}{2}} \label{dkw}.
\end{align} 
The constant $D_k^{*}(w)$ represents the strength of the ``super charge" in the saddle point density expression in Eq.~\eqref{dens-s:klm1}. Note from Eq.~\eqref{dens-s:klm1} that $\bar{l}_k(w)$ is the position of the right edge of the extended part of the density. Since $\bar{l}_k(w)<w$, as can be seen from Eq.~\eqref{barl}, there is a region $\bar{l}_k(w)<y<w$ devoid of particles. A schematic representation of the density is given in Fig.~\ref{fig:wdens}c. Interestingly, as the wall is pushed further to the left, more and more particles accumulate at the wall and at the critical wall position $w_c(k)$ the strength of the delta function becomes one, the systems undergoes a first-order transition~\cite{kethepalli2021harmonically}.
The value of $l_{k}(w)$ is determined from the normalisation condition which can be expressed in terms of the auxiliary function
\begin{equation}\label{gkwlm1}
h_k(w) = \frac{w+l_k(w)}{\bar{l}_k(w)+l_k(w)}, \text{  for  } -2<k<-1,
\end{equation}
as
\begin{equation}\label{normklm1}
 \left(1 + (h_k(w)-1)\frac{2(k+2)}{(k+1)}\right)~_2F_1[1,-(k+2),-\frac{k+1}{2}, 1-h_k(w)]^{-\frac{1}{k+2}} = \frac{w}{l_k^{\rm uc}}.
\end{equation}
The details of the computation leading to Eqs.~\eqref{gkwlm1} and~\eqref{normklm1} are given in~\ref{apndklm1}.

The scaled energy functional $\tilde{\mathcal{E}}_k[\rho^*_k]$ is obtained by substituting the density profile $\rho^*_k(y, w)$ in Eq.~\eqref{scaled-field}. Using this energy functional in Eq.~\eqref{LDF-general}, we compute the LDF $\Phi_{\pm}(w, k)$ which is given by (see~\ref{apndklm1} for details)
\begin{equation}
\begin{split}
\Phi_{-}(w, k) &= (l_k^{\rm uc})^2 ~_2F_1[1, -(k+2), -\frac{k+1}{2}, 1-h_k(w)]^{-\frac{2}{k+2}} \Bigg[\frac{1}{4 k}- \frac{(h_k(w)-1)}{k(k+1)}\\&- (k+2)\frac{(h_k(w)-1)^2}{k(k+1)^2} +  n_k(w)^2  \\&-n_k(w) \frac{~_2F_1[1, -(k+3), -\frac{k+1}{2}, 1-h_k(w)]}{~_2F_1[1, -(k+2), -\frac{k+1}{2}, 1-h_k(w)]}   \\&+\frac{k+5}{4(k+4)} \frac{~_2F_1[1, -(k+4), -\frac{k+1}{2}, 1-h_k(w)]}{~_2F_1[1, -(k+2), -\frac{k+1}{2}, 1-h_k(w)]}\Bigg] -(l_k^{\rm uc})^2\frac{k+2}{2 k (k+4)}\label{ldf:lklm1_2},
\end{split}
\end{equation}
where,
\begin{equation}\label{qkwlm1}
n_k(w)= \frac{l_k(w)}{\tilde{L}_k(w)} = \frac{1}{2} + \frac{1-h_k(w)}{k+1}.
\end{equation}
The calculation for the right LDF is exactly the same as in Section.~\ref{sec:regime2}, hence the expression is the same as Eq.~\eqref{ldf:rkgm1_3} i.e.
\begin{align}\label{ldf:rklm1_3}
%\begin{split}
\Phi_{+}(w, k) = &\left(l_k^{\rm uc}\right)^2 \frac{32 (h_k^{\rm uc}(w)^{-1}-1)^{\frac{3-k}{2}}B(2+k, \frac{5-k}{2})}{(k+3)(k+5)(k+7)} \notag \\ & \quad  \quad \times ~_2F_1 \big[-\frac{k+1}{2},\frac{k+3}{2}, \frac{5-k}{2}, 1-h_k^{\rm uc}(w)^{-1}\big],
%\end{split}
\end{align}
where $h_k^{\rm uc}(w) = g_k^{\rm uc}(w) = 2 l_k^{\rm uc}/(w + l_k^{\rm uc})$. Once again these LDF are verified numerically using importance sampling method in Fig.~\ref{fig:km1p5} which demonstrates an excellent agreement.

The asymptotic behaviour of $\Phi_{-}(w, k)$ is obtained by performing the series expansion about $h_k(w)=1$, namely
\begin{equation}\label{klm1:epsi}
\begin{split}
h_k(w) &= 1-\frac{k+1}{k+3}\frac{l_k^{\rm uc}-w}{2l_k^{\rm uc}}+o(l_k^{\rm uc}-w).
\end{split}
\end{equation}
Sbstituting this asymptotic behaviour in Eq.~\eqref{ldf:lklm1_2} on finds
\begin{equation}\label{phimekmklm1}
\Phi_-(w, k) \approx \frac{2(k+2)(k+5)}{3(3+k)^2(k-1)(k-3)l_k^{\rm uc}}\left(l_k^{\rm uc}-w \right)^{e^-_k} \text{   with   } e^-_k = 3.
\end{equation}
Hence in this regime also the system undergoes a $3^{\rm rd}$ order pulled to pushed phase transition.  Finally, the asymptotic behaviour of the right LDF is given by Eq.~\eqref{asym:rdf} namely 
\begin{equation}\label{phipekpklm1}
\Phi_+(w, k) \approx \frac{2 (1-k) B\left(k+2,\frac{5-k}{2}\right)(l_k^{\rm uc})^{-e^+_k}}{3 A_k (k+5) (k+7) \left| k\right|  B\left(\frac{k+3}{2},1-k\right)} \left(w- l_k^{\rm uc}\right)^{e^+_k}  \text{   with   } e^+_k = \frac{3-k}{2}.
\end{equation}
%replaced large deviation functions with LDF
\section{Discussions and Conclusions}
\label{conclusion}
In this paper, we investigated the fluctuations of the position of the rightmost (edge) particle $y_{\max} = x_{\max}/L_N$ of harmonically confined Riesz gas [Eq.~\eqref{microhami}]. We studied both typical and the atypical fluctuations of $y_{\max}$ separately. From numerical analysis  we found that the typical fluctuations characterised by the variance scales as $N^{-2 \eta_k}$ with $N$. Similar to the scaling of the support, the exponent $\eta_k$ associated with the variance of $y_{\max}$, also depends on $k$ non-monotonically as shown in Fig.~\ref{fig:etakvsk}. We have provided a physical understanding of the $k$ dependence of $\eta_k$ based on Hessian theory and a scaling argument. For $-2<k<0$, the assumption that the full distribution of $y_{\max}$ has a single length scale led us to conjecture an explicit expression of $\eta_k$ given in Eq.~\eqref{conjecture}. This conjecture was tested against the MC simulations in Fig.~\ref{fig:etakvsk} and we found remarkable agreement. For $k>0$, we found that the exponent $\eta_k$ matches extremely well with the one computed from the Hessian theory [see Fig.~\ref{fig:etakvsk}]. For all $k$, we found that the distribution of $y_{\max}$ when shifted by mean and scaled by the $\sigma_{y_{\max}} \sim N^{-\eta_k}$, exhibits a remarkable data collapse leading to a scaling distribution which is non-Gaussian in general.

The atypical fluctuations to the left and right of the mean are described by the left and the right LDF.
We computed the explicit expressions for these LDF in different regimes of $k$. We found that their asymptotic behaviour near the edge of the unconstrained density are $k$ dependent and shown in Table.~\ref{table:epmk} and Fig.~\ref{fig:ekg1}. This difference is a consequence of the different mechanisms by which the saddle point density of the gas gets modified in the presence of a wall. 
A manifestation of this difference in the asymptotic behaviour of the LDF is demonstrated in terms of the analytic properties of appropriately defined free energies which exhibits the third-order pulled to pushed phase transition $ \forall k>-2$. Therefore our results reveal a striking universality of the third-order phase transition in family of models that fall outside the paradigm of Coulomb systems and RMT.  All our results hold for temperature $T< L_N^2$. 

A straightforward extension of our results on LDF and phase transition can be made for other confining potential of the form $|x|^{\delta}$ with $\delta>0$ and $k>-\delta$. Fascinating open and challenging questions include (a) deriving the explicit expression of $\eta_k$ (b) finding the explicit form of the distribution $\mathcal{F}_{\beta}^{(k)}(z)$ for the typical fluctuations and (c) exploring the possibility of multiple scales in the fluctuations.  

\ack
We thank S. Santra for sharing the ground state configuration data obtained using the BFGS algorithm. We thank A. Dhar for helpful discussions and suggestions. M.K. would like to acknowledge support from the Project 6004-1 of the Indo-French Centre for the Promotion of Advanced Research (IFCPAR), Ramanujan Fellowship (SB/S2/RJN114/2016), SERB Early Career Research Award (ECR/2018/002085) and SERB Matrics Grant (MTR/2019/001101) from the Science and Engineering Research Board (SERB), Department of Science and Technology, Government of India. This research was supported in part by the International Centre for Theoretical Sciences (ICTS) for enabling discussions during the program - Fluctuations in Nonequilibrium Systems: Theory and applications (Code:ICTS/Prog- fnsta2020/03). J.K., M.K. and A.K. acknowledge support from the Department of Atomic Energy, Government of India, under Project No. RTI4001. A.K. acknowledges support from DST, Government of India grant under Project No. ECR/2017/000634. S.N.M. thanks the warm hospitality of the Weizmann Institute as a visiting Weston fellow and ICTS where this work was completed. D.M. acknowledges the support of the Center of Scientific Excellence at the Weizmann Institute of Science.

\appendix
\addtocontents{toc}{\fixappendix}

\section{System size dependence of $\sigma_{y_{\max}}$}
\label{apndtyp}
\begin{figure}[h!]
    \centering
    \includegraphics[scale=0.5]{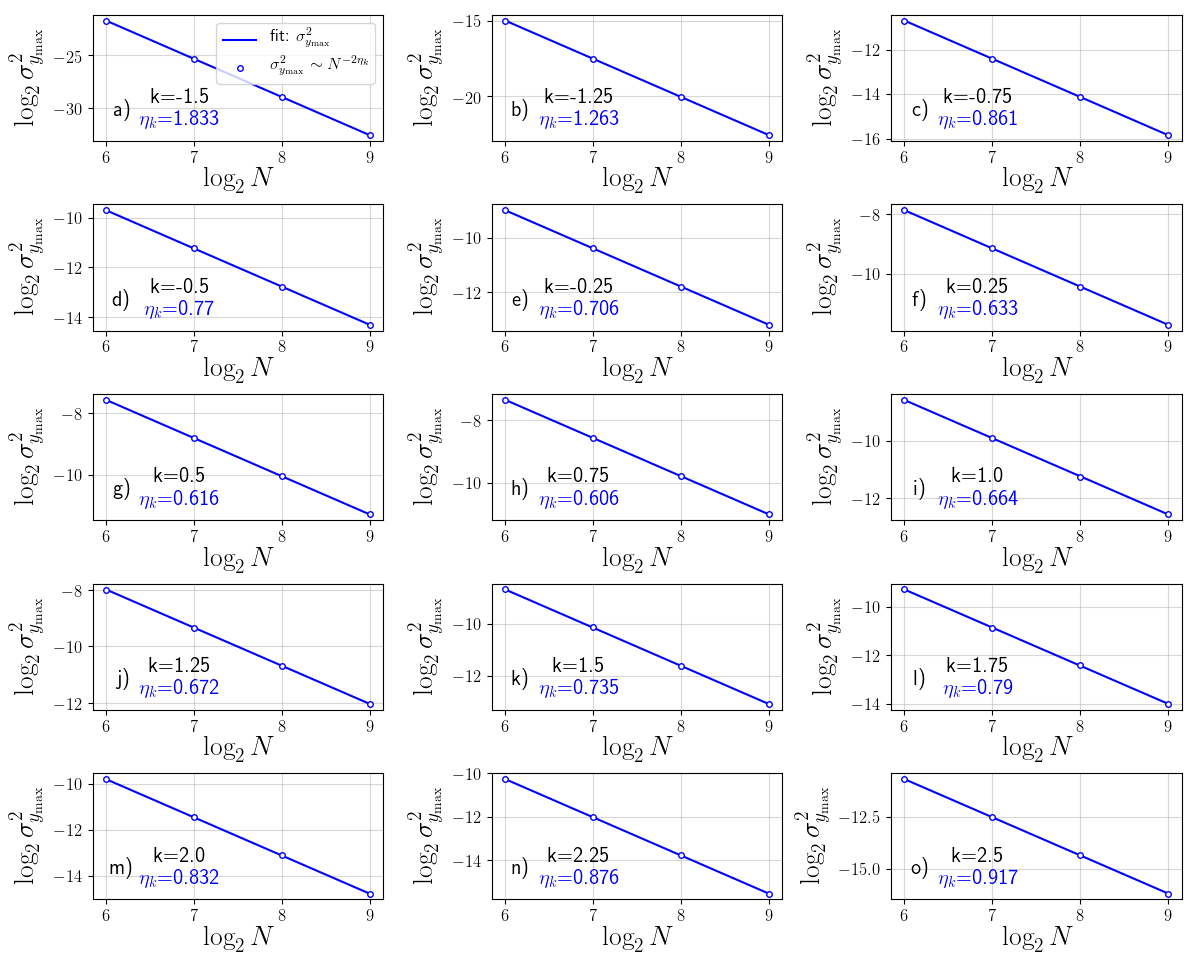}
    \caption{Plots of the logarithm of the variance of the position of the rightmost particle obtained using the MC simulations of the confined Riesz gas Hamiltonian [Eq.~\eqref{microhami}] versus the logarithm (base-$2$) of system size $N$ (disks) for (a) $k=-1.5$, (b) $k=-1.25$, (c) $k=-0.75$, (d) $k=-0.5$, (e) $k=-0.25$, (f) $k \to 0$, (g) $k=0.25$, (h) $k=0.5$, (i) $k=0.75$, (j) $k=1.0$, (l) $k=1.25$, (n) $k=1.5$, (m) $k=1.75$,  (n) $k=2.25$, (o) $k=2.5$. Here we use a linear fit (solid lines) of the data to extract the slope $-2\eta_k$.}
    \label{fig:etakall}
\end{figure}

\begin{figure}[h!]
    \centering
    \includegraphics[scale=0.5]{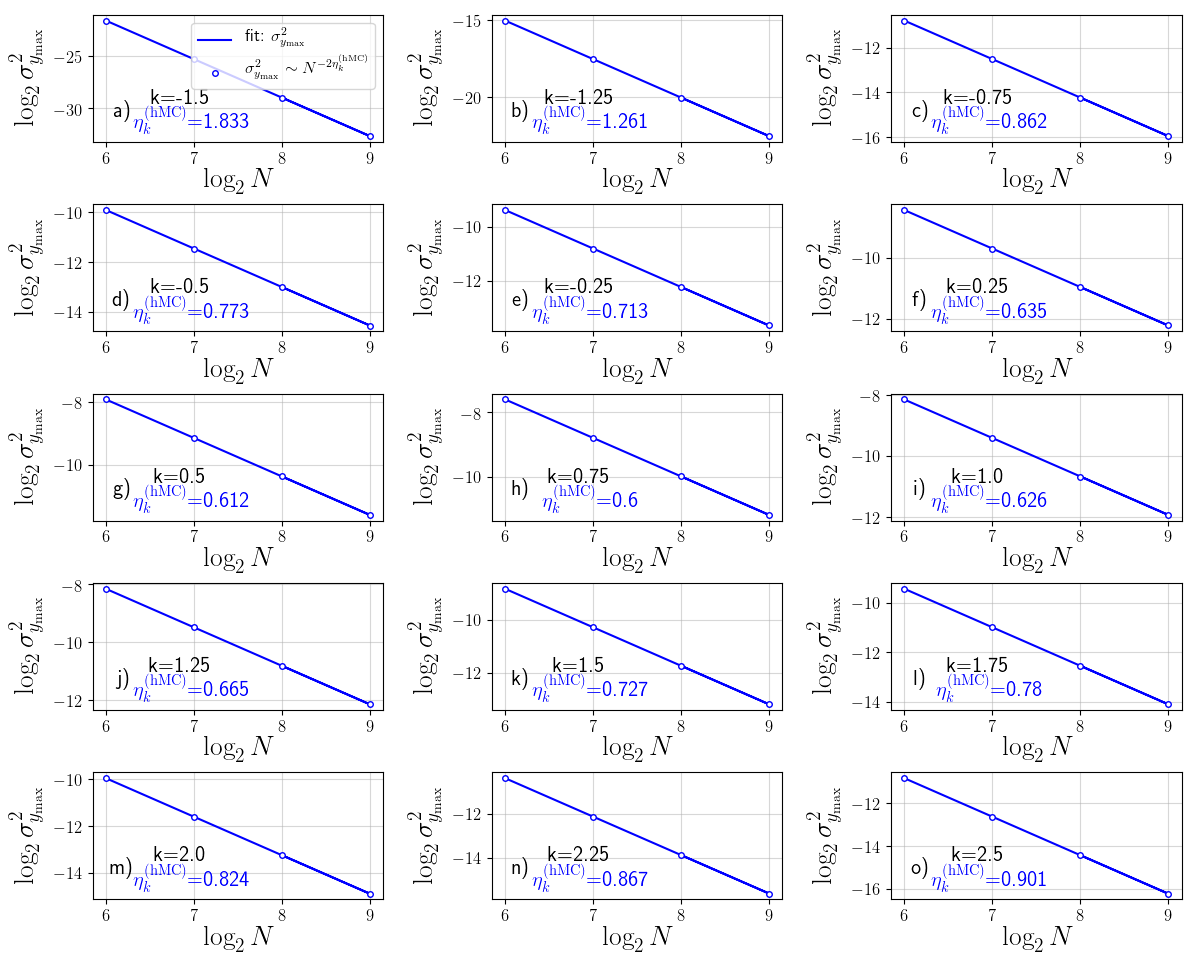}
    \caption{Plots of the logarithm of the variance of the position of the rightmost particle obtained using the MC simulations for the Hessian Hamiltonian [Eq.~\eqref{hess:hami}] versus the logarithm (base-$2$) of system size $N$ (disks) for (a) $k=-1.5$, (b) $k=-1.25$, (c) $k=-0.75$, (d) $k=-0.5$, (e) $k=-0.25$, (f) $k \to 0$, (g) $k=0.25$, (h) $k=0.5$, (i) $k=0.75$, (j) $k=1.0$, (l) $k=1.25$, (n) $k=1.5$, (m) $k=1.75$,  (n) $k=2.25$, (o) $k=2.5$. Here we use a linear fit (solid lines) of the data to extract the slope $-2\eta_k^{\rm (hMC)}$.}
    \label{fig:etakall_h}
\end{figure}

Here, we provide plots of the numerical data for $\log_2 \sigma^2_{y_{\max}}$ vs $\log_2 N$ (symbols) in Fig.~\ref{fig:etakall} (for confined Riesz gas [Eq.~\eqref{microhami}]) and Fig.~\ref{fig:etakall_h} (for the Hessian Hamiltonian [Eq.~\eqref{hess:hami}]) for different values of $k$. For each $k$ the slope of the linear fit (solid lines) of this data provides the exponent ($\eta_k, \eta_k^{\rm (hMC)}$) which we plot in Fig.~\ref{fig:etakvsk}.

\section{Large deviation functions}
\label{apndixa}
In this appendix, we present the details of the derivation of the LDF ($\Phi_{\pm}(w, k)$) which characterise the distribution of the atypical fluctuations. To obtain the LDF $\Phi_{\pm}(w, k)$, we use the energy functional Eq.~\eqref{scaled-field} in the formal expression of the LDF given in Eq.~\eqref{LDF-general}. We study all the three regimes of $k$ separately and compute $\Phi_{\pm}(w, k)$, for (a) $k>1$ given in Eq.~\eqref{ldf:lkg1_1} and~\eqref{ldf:rkg1_1}, (b)  $-1<k<1$  given in Eq.~\eqref{ldf:lkgm1_2} and~\eqref{ldf:rkgm1_3} and (c) $-2<k\leq-1$ given in Eq.~\eqref{ldf:lklm1_2} and~\eqref{ldf:rklm1_3}. 

\subsection{Regime $1: k>1$}
\label{apndkg1}

\noindent \textit{Left large deviation function:} We start by rewriting the expression of the left LDF $\Phi_{-}(w, k)$ given in Eq.~\eqref{LDF-general} as
\begin{equation}\label{LDF:left_2}
\begin{split}
\Phi_{-}(w, k) =& \frac{1}{2(k+1)}\left(l_k(w)^2 - (l_k^{\rm uc})^2+k \int_{-l_k(w)}^w dy\, y^2 \rho^*_k(y,w)-k \int_{-l_k^{\rm uc}}^{l_k^{\rm uc}} dy\, y^2 \rho_{k,\rm uc}^*(y)\right).
\end{split}
\end{equation}
Substituting the expression of the constrained density from Eq.~\eqref{kg1densii}, unconstrained density from Eq.~\eqref{rho_uc} in Eq.~\eqref{LDF:left_2}, and using a variable transformation 
\begin{equation}
 z = \frac{y+l_k(w)}{2l_k(w)},
\end{equation}
the left LDF can be expressed in terms of an auxiliary variable, $m_k(w) = \frac{w+ l_k(w)}{2 l_{k}(w)}$ [see Eq.~\eqref{mkw}] as
\begin{equation}
\begin{split}
\Phi_{-}(w, k) =& \frac{1}{2(k+1)}\Bigg[ l_k(w)^2 - (l_k^{\rm uc})^2+k A_k (l_k(w))^{2 \gamma_k+3} \int_{0}^{m_k(w)} dz\, \left(z-\frac{1}{2}\right)^2 \left(z(1-z)\right)^{\gamma_k}\\&-k A_k (l_k^{\rm uc})^{2 \gamma_k+3} \int_{0}^{1} dz\, \left(z-\frac{1}{2}\right)^2 (z(1-z))^{\gamma_k} \Bigg].
\end{split}
\end{equation}
Using Eq.~\eqref{l_{k}(w)-k>1} in Eq.~\eqref{LDF:left_2} we can further simplify the expression in terms of incomplete Beta function as given in Eq.~\eqref{ldf:lkg1_1}, i.e.,
\begin{equation}
\begin{split}
\Phi_{-}(w, k) &= \frac{l_k(w)^2}{2(k+1)} \Bigg[1+ \frac{4 k}{B(m_k(w),\gamma_k+1,\gamma_k+1)} \bigg(B(m_k(w), \gamma_k+3, \gamma_k+1)\\&-B(m_k(w), \gamma_k+2, \gamma_k+1)+ \frac{B(m_k(w), \gamma_k+1, \gamma_k+1)}{4}\bigg)\Bigg]-\frac{(l_k^{\rm uc})^2(k+2)}{2(3k+2)}.
\end{split}
\end{equation}
\noindent \textit{Right large deviation function:} The formal expression for the PDF of the atypical fluctuation to the right of the mean is given in Eq.~\eqref{PDF:right} from which the right LDF can be written as
\begin{equation}\label{LDF:right}
\begin{split}
\Phi_+(w, k)= \Bigg(\frac{w^2-{l_k^{\rm uc}}^2}{2}  \left. \right. +\frac{J N}{L_N^{k+2}} \int dy\rho_{k,{\rm uc}}^{\rm *}(y) \left( \frac{1}{|w-y|^{k}} - \frac{1}{|l_k^{\rm uc}-y|^{k}}\right) \Bigg).
\end{split}
\end{equation}
Note that the right LDF has two parts: 1) an external potential term and 2) an interaction term. In the large-$N$ limit the interaction term becomes negligible, since $N \to \infty$ implies $N/L_N^{k+2} \to 0$, compared to the external potential term, which is a consequence of the short-ranged nature of the force. Hence, we obtain Eq.~\eqref{ldf:rkg1_1} which is given by
\begin{equation}
\Phi_{+}(w, k) = \frac{w^2 - \left(l_k^{\rm uc}\right)^2}{2}.
\end{equation}

\subsection{Regime $2: -1<k<1$}
\label{apndkl1}
\noindent{\it Left large deviation function}: The expression for left LDF given in Eq.~\eqref{LDF-general} is simplified using the formal equation for the chemical potential Eq.~\eqref{kl1 chem eq} which gives
\begin{equation}\label{ldf:lkgm1_1}
\begin{split}
\Phi_{-}(w, k) = \int_{-l_k(w)}^w dy\, \frac{y^2}{4}  \rho^*_k(y, w) -  \int_{-l_k^{\rm uc}}^{l_k^{\rm uc}} dy\, \frac{y^2}{4} \rho_{k, \rm uc}^*(y) + \frac{2\mu_k^*(w)-(l_k^{\rm uc})^2}{4}.
\end{split}
\end{equation}
We substitute the expression for the constrained density profile [$\rho^*_k(y, w)$] from Eq.~\eqref{kl1dens} and the unconstrained density profile [$\rho_{k, \rm uc}^*(y)$] from Eq.~\eqref{rho_uc} in the above simplified expression Eq.~\eqref{ldf:lkgm1_1}. Using the change of variable
\begin{equation}\label{vart:kgm1}
 z = \frac{y+l_k(w)}{L_k(w)} \text{  with  } L_k(w) = w + l_k(w),
\end{equation}
where $L_k(w)$ is the total size of the support, one can express Eq.~\eqref{ldf:lkgm1_1} in terms of auxiliary variables
\begin{equation}\label{aux:kgm1}
g_k(w) = \frac{\tilde{l}_k(w) +l_k(w)}{L_k(w)} \text{ and } q_k(w) = \frac{l_k(w)}{L_k(w)} = \frac{1}{2} + \frac{g_k(w)-1}{k+1}.
\end{equation}
In terms of these auxilary variables, the left LDF takes the form
\begin{equation}
\begin{split}
\Phi_{-}(w, k) &= A_k \left(L_k(w)\right)^{2 \gamma_k+3}\int_{0}^1 dz\, \frac{\left(z-q_k(w)\right)^2}{4} (g_k(w)-z) z^{\frac{k+1}{2}} \left(1-z\right)^{\frac{k-1}{2}} \\&-  A_k \left(2l_k^{\rm uc}\right)^{2 \gamma_k+3}\int_{0}^1 dz\, \frac{\left(z-\frac{1}{2}\right)^2}{4} \left(z(1-z)\right)^{\frac{k+1}{2}} + \frac{2\mu_k^*(w)-(l_k^{\rm uc})^2}{4} \label{ldf:lkgm1_1.5}.
\end{split}
\end{equation}
To proceed further we now need to compute the chemical potential $\mu_k^*(w)$.

\begin{figure}[h!]
    \centering
    \includegraphics[scale=0.7]{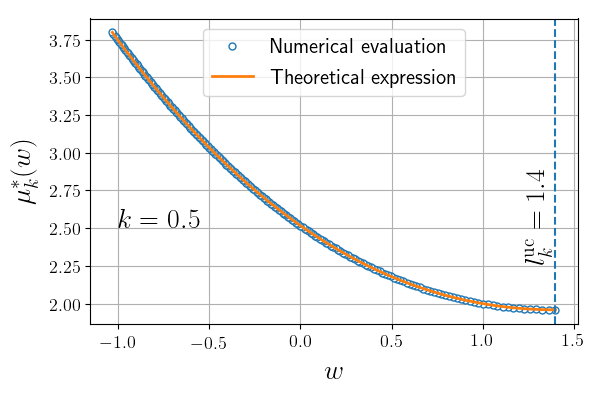}
    \caption{\textit{Regime 2 ($-1<k < 1$):} For $k=0.5$, the plot of chemical potential as a function of the wall position, computed from numerical integration (circles) of Eq.~\eqref{akgm1:chem1} for $z=0$. This evaluation is compared with the simplified expression given in Eq.~\eqref{kl1 chem eq-1} (solid lines). The parameter used for the computations is $J=1$.}
    \label{fig:kl1chem}
\end{figure}

The chemical potential in Eq.~\eqref{kl1 chem eq} can be simplified by a variable transformation given in Eq.~\eqref{vart:kgm1} which gives
\begin{equation}\label{akgm1:chem1}
\mu_k^*(w) = L_k(w)^2 \Bigg[\frac{(z-q_k(w))^2}{2}  + A_k {\rm sgn}(k) \int_0^1 dr \frac{r^{\frac{k+1}{2}} (1-r)^{\frac{k-1}{2}}}{|z-r|^k} (g_k(w)-r)\Bigg].
\end{equation}
The integral in the square bracket  can be spilt into two integrals as 
\begin{equation}\label{akgm1:chemintegral}
\begin{split}
\int_0^1 dr \frac{r^{\frac{k+1}{2}} (1-r)^{\frac{k-1}{2}}}{|z-r|^k} (g_k(w)-r) =& \int_0^z dr \frac{r^{\frac{k+1}{2}} (1-r)^{\frac{k-1}{2}}}{(z-r)^k} (g_k(w)-r) \\&+ \int_z^1 dr \frac{r^{\frac{k+1}{2}} (1-r)^{\frac{k-1}{2}}}{(r-z)^k} (g_k(w)-r).
\end{split}
\end{equation}
The first integral in Eq.~\eqref{akgm1:chemintegral} can be further simplified by using the following variable transformation
\begin{equation}
s_1 = \frac{z-r}{z},
\end{equation}
which gives
\begin{equation}\label{akgm1:chemintegral1}
\begin{split}
\int_0^z dr & \frac{r^{\frac{k+1}{2}} (1-r)^{\frac{k-1}{2}}}{(z-r)^k} (g_k(w)-r)\\&=z^2 \int_0^1 d s_1 \frac{s_1^{-k} (1-s_1)^{\frac{k+1}{2}}}{\left(s_1+\frac{1-z}{z}\right)^{\frac{1-k}{2}}}\left(s_1+\frac{g_k(w)-z}{z}\right)\\
&= \left(\frac{1-z}{z}\right)^{\frac{k-1}{2}}\Bigg[z^2  B\left(2-k,\frac{k+3}{2}\right)  ~_2F_1\left[\frac{1-k}{2},2-k,\frac{7-k}{2},\frac{z}{z-1}\right] \\&+ z(g_k(w)-z)B\left(1-k,\frac{k+3}{2}\right)  ~_2F_1\left[\frac{1-k}{2},1-k,\frac{5-k}{2},\frac{z}{z-1}\right]\Bigg].
\end{split}
\end{equation}
%here $~_2F_1[a,b,c,u] = B(b, c-b)^{-1}\int_0^1 ds \frac{s^{b-1}(1-s)^{c-1-b}}{(1-us)^a}$ is the hypergeometric function.
Similarly the second integral in Eq.~\eqref{akgm1:chemintegral} can be further simplified by using the following variable transformation
\begin{equation}
s_2 = \frac{r-z}{1-z},
\end{equation}
which gives
\begin{equation}\label{akgm1:chemintegral2}
\begin{split}
\int_z^1 dr & \frac{r^{\frac{k+1}{2}} (1-r)^{\frac{k-1}{2}}}{(r-z)^k} (g-r) \\&=(1-z)^2 \int_0^1 d s_2\frac{s_2^{-k} (1-s_2)^{\frac{k-1}{2}}}{\left(s_2+\frac{z}{1-z}\right)^{-\frac{1+k}{2}}}\left(\frac{g-z}{1-z}-s_2\right)\\
&= \left(\frac{z}{1-z}\right)^{\frac{k+1}{2}} \Bigg[(1-z) (g-z) B\left(1-k,\frac{k+1}{2}\right)  ~_2F_1\left[-\frac{k+1}{2},1-k,\frac{3-k}{2},\frac{z-1}{z}\right] \\&-(1-z)^2 B\left(2-k,\frac{k+1}{2}\right) ~_2F_1\left[-\frac{k+1}{2},2-k,\frac{5-k}{2},\frac{z-1}{z}\right]\Bigg] .
\end{split}
\end{equation}
Note that the argument of the hypergeometric function in Eq.~\eqref{akgm1:chemintegral1} is $\frac{z}{z-1}$ whereas in Eq.~\eqref{akgm1:chemintegral2} it is  $\frac{z-1}{z}$. Hence to simplify this further we use the following hypergeometric function identity~\cite{hyper}
\begin{equation}\label{identity:hype}
\begin{split}
B(b+1,a+1) &~_2F_1\left[a+1,c,a+b+2,-\frac{1}{u}\right] \\& = u^{a+1} B(-a+c-1,a+1) ~_2F_1\big[a+1,-b,a-c+2,-u\big] \\& + u^c B(b+1,a-c+1) ~_2F_1\big[c,-a-b+c-1,c-a,-u\big].
\end{split}
\end{equation}
Substituting the above identity in Eq.~\eqref{akgm1:chemintegral1}, we can express the Eq.~\eqref{akgm1:chemintegral1} as a function of $\frac{z-1}{z}$ instead of $\frac{z}{z-1}$. Then using the new expression in terms of $\frac{z-1}{z}$ and Eq.~\eqref{akgm1:chemintegral2} in Eq.~\eqref{akgm1:chem1}, we can simplify the expression of $\mu_k^*(w)$ which after a tedious calculation gives
\begin{equation}\label{kl1 chem eq-1}
\mu_k^*(w) = L_k(w)^2 \left(\frac{1}{8 k}+\frac{(g_k(w)-1)^2}{2 (k+1)^2}+\frac{g_k(w)-1}{2 k (k+1)}\right).
\end{equation}
Since the chemical potential is a constant, we can independently find the value of $\mu_k^*(w)$ by substituting $z=0$ or $z=1$ in Eq.~\eqref{akgm1:chem1}. This expression is verified by numerically evaluating the integrals in Eq.~\eqref{akgm1:chem1} directly
and comparing with Eq.~\eqref{kl1 chem eq-1}. This comparision is shown in Fig.~\ref{fig:kl1chem}.
The remaining integrals in the left LDF given in Eq.~\eqref{ldf:lkgm1_1.5} can be expressed as beta functions. Now using the expression of the chemical potential from Eq.~\eqref{kl1 chem eq-1} in Eq.~\eqref{ldf:lkgm1_1.5} we get 
\begin{equation}
\begin{split}
\Phi_{-}(w, k) = & (k+2)(l_k^{\rm uc})^2\Bigg[ \left(1+\frac{2(k+2)}{k+1}(g_k(w)-1)\right)^{-\frac{k+4}{k+2}} \Big[\frac{1}{2k(k+4)}\\&+\frac{(g_k(w)-1)}{k(k+1)}+\frac{2(g_k(w)-1)^2}{k(k+1)^2} + \frac{4(g_k(w)-1)^3}{(1+k)^3}\Big] - \frac{1}{2k(k+4)}\Bigg],
\end{split}
\end{equation}
as announced in Eq.~\eqref{ldf:lkgm1_2}. Here, the length of the constrained support $L_k(w) = w/(1-q_k(w))$ was expressed in terms of the auxilary variable $g_k(w)$ which was obtained using Eq.~\eqref{l_{k}(w)-k<1}.

\noindent{\it Right large deviation function}: In this regime both the external potential term and the interaction term  contribute equally to the right LDF in Eq.~\eqref{LDF:right} since $N \to \infty$ implies $ N/L_N^{k+2} \to 1$. Hence the right LDF is given by
\begin{equation}
\begin{split}
\Phi_{+}(w, k) = & A_k {\rm sgn}(k)\left(\int_{-l_k^{\rm uc}}^{l_k^{\rm uc}} dy \frac{\left(\left(l_k^{\rm uc}\right)^2 - y^2\right)^{\frac{k+1}{2}}}{(w-y)^{k}}\right)+\frac{w^2}{2}-\frac{(l_k^{\rm uc})^2}{2k}
\label{ldf:rkgm1_2}.
\end{split}
\end{equation}
Using the following variable transformation
\begin{equation}
z= \frac{l_k^{\rm uc}-y}{2 l_k^{\rm uc}} \text{  and  } g_k^{\rm uc}(w) ^{-1} = \frac{w+l_k^{\rm uc}}{2 l_k^{\rm uc}},
\end{equation} 
Eq.~\eqref{ldf:rkgm1_2} can be expressed as
\begin{equation}
\begin{split}
\Phi_{+}(w, k) = &\left(2 l_k^{\rm uc}\right)^2 \left( A_k {\rm sgn}(k) \int_{0}^{1} dz \frac{\left(z(1-z)\right)^{\frac{k+1}{2}}}{(z+g_k^{\rm uc}(w) ^{-1} -1)^{k}}+\frac{\left(2 g_k^{\rm uc}(w)^{-1}-1\right)^2}{8}-\frac{1}{8k}\right)
\label{ldf:rkgm1_2.5}.
\end{split}
\end{equation}
This can be further simplified to
\begin{equation}
\begin{split}
\Phi_{+}(w, k) = &\left(l_k^{\rm uc}\right)^2 \frac{32 (g_k^{\rm uc}(w)^{-1}-1)^{\frac{3-k}{2}}B(2+k, \frac{5-k}{2})}{(k+3)(k+5)(k+7)} \times \\&~_2F_1[-\frac{k+1}{2},\frac{k+3}{2}, \frac{5-k}{2}, 1-g_k^{\rm uc}(w)^{-1}].
\end{split}
\end{equation}
which is Eq.~\eqref{ldf:rkgm1_3} of the main text. 
\subsection{Regime $3: -2<k\leq-1$}
\label{apndklm1}

\noindent
We start by expressing the normalization condition $\int_{-l_k(w)}^{w} dy \rho_k^*(y, w) = 1$ in terms of the auxiliary variable $h_k(w)$ given in Eq.~\eqref{gkwlm1}. Using the expression of the constrained density given in Eq.~\eqref{dens-s:klm1} and the variable transformation 
\begin{equation}\label{vart:klm1}
 z = \frac{\bar{l}_k(w)-y}{\tilde{L}_k(w)} \text{  with  } \tilde{L}_k(w) = \bar{l}_k(w) + l_k(w),
\end{equation}
we get 
\begin{equation}\label{aklm1:nom1}
\begin{split}
\int_{-l_k(w)}^{w} \rho_k^*(y, w) &= \left(\tilde{L}_k(w)\right)^{2 \gamma_k+1} \Bigg( A_k \int_{0}^1 dz\, \frac{z^{\frac{k+3}{2}} (1-z)^{\frac{k+1}{2}} }{z+h_k(w)-1} + \frac{h_k(w)^{\frac{k+1}{2}}(h_k(w)-1)^{\frac{k+3}{2}}}{k(k+1)}\Bigg) \\&
= \left(\tilde{L}_k(w)\right)^{2 \gamma_k+1} \Bigg( A_k \Big((h_k(w)-1)^{\frac{k+3}{2}} h_k(w)^{\frac{k+1}{2}} B\left(-\frac{k+3}{2},\frac{k+5}{2}\right)\\&+B\left(\frac{k+3}{2},\frac{k+3}{2}\right) ~_2F_1[1,-(k+2),-\frac{k+1}{2},1-h_k(w)]\Big) \\&+ \frac{h_k(w)^{\frac{k+1}{2}}(h_k(w)-1)^{\frac{k+3}{2}}}{k(k+1)}\Bigg),
\end{split}
\end{equation}
where $\tilde{L}_k(w)$ is the size of the support of the extended part of the density profile.
This equation can be further simplified by using the fact that
\begin{equation}
A_k B\left(-\frac{k+3}{2}, \frac{k+5}{2}\right) = \frac{1}{|k|(k+1)},
\end{equation}
which then gives
\begin{equation}\label{aklm1:nom2}
\begin{split}
\left(\frac{\tilde{L}_k(w)}{2 l_k^{\rm uc}}\right)^{2 \gamma_k+1}~_2F_1[1,-(k+2),-\frac{k+1}{2},1-h_k(w)] =1.
\end{split}
\end{equation}
The position of the wall can be expressed as $w = \tilde{L}_k(w) \big(h_k(w)-n_k(w) \big)$ where 
\begin{equation}\label{qkwlm1}
n_k(w)= \frac{l_k(w)}{\tilde{L}_k(w)} = \frac{1}{2} + \frac{1-h_k(w)}{k+1}.
\end{equation}
The relation between $n_k(w)$ and $h_k(w)$ is obtained using Eq.~\eqref{barl}. This finally gives Eq.~\eqref{normklm1}.

\begin{figure}[h!]
    \centering
    \includegraphics[scale=0.7]{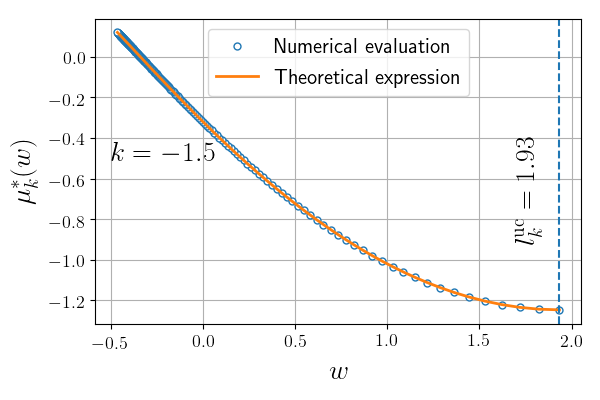}
    \caption{\textit{Regime 3 ($-2<k < -1$):} For $k=-1.5$, the plot of chemical potential as a function of wall position, computed from numerical integration (circle) given in Eq.~\eqref{klm1 chem eq} and the simplified expression (solid lines) given in Eq.~\eqref{klm1 chem eq-1}. Here we considered $J=1$.}
    \label{fig:klm1chem}
\end{figure}

\noindent{\it Left Large deviation function}: Since the structure of the field theory is the same, the formal expression of the left LDF in this regime is similar to the previous regime given in Eq.~\eqref{ldf:lkgm1_1}. However, the density profiles here are different, as  in this regime the constrained density given in Eq.~\eqref{dens-s:klm1} has a delta function of strength $D^*_k(w)$ and a disjoint extended part denoted as $\rho^*_b(y, w)$. For convenience we here write the explicit expression of $\rho^*_b(y, w)$ from Eq.~\eqref{dens-s:klm1} as 
\begin{equation}\label{dens:bulk}
\rho^*_b(y, w)= A_k \frac{\left(l_{k}(w)+y\right)^{\frac{k+1}{2}} \left(\bar{l}_k(w)-y\right)^{\frac{k+3}{2}}}{(w-y)}.
\end{equation}
We use $\rho^*_k(y,w) = \rho^*_b(y,w) \mathbb{I}[-l_k(w)<y \leq \bar{l}_k(w)]+ D^*_k(w) \delta(y-w)$ in Eq.~\eqref{ldf:lkgm1_1} which gives
\begin{equation}
\begin{split}
\Phi_{-}(w, k) =  &\frac{2\mu_k^*(w)+ D_k(w) w^2}{4} +\int_{-l_k(w)}^{\bar{l}_k(w)} dy \frac{y^2}{4}\rho^*_b(y, w)\\&-\frac{\left(l_k^{\rm uc}\right)^2}{4} -\int_{-l_k^{\rm uc}}^{l_k^{\rm uc}} dy \frac{y^2}{4}\rho^*_{k, \rm uc}(y)\label{ldf:lklm1_1}.
\end{split}
\end{equation}
We substitute the expressions of $\rho^*_b(y, w)$ from Eq.~\eqref{dens:bulk}, $D^*_k(w)$ from Eq.~\eqref{dkw} and $\rho^*_{k, \rm uc}(y)$ from Eq.~\eqref{rho_uc} in the above equation. Using the variable change given in Eq.~\eqref{vart:klm1}, we can now express the left LDF in terms of auxiliary variables $h_k(w)$ and $n_k(w)$ given in Eq.~\eqref{gkwlm1} and Eq.~\eqref{qkwlm1}, respectively. The left LDF then becomes
\begin{equation}
\begin{split}
\Phi_{-}(w, k) &= A_k \left(\tilde{L}_k(w)\right)^{2 \gamma_k+3}\int_{0}^1 dz\, \frac{\left(1-q_k(w)-z\right)^2}{4}  \frac{z^{\frac{k+3}{2}} \left(1-z\right)^{\frac{k+1}{2}}}{z+h_k(w)-1} \\&-  A_k \left(2l_k^{\rm uc}\right)^{2 \gamma_k+3}\int_{0}^1 dz\, \frac{\left(z-\frac{1}{2}\right)^2}{4} \left(z(1-z)\right)^{\frac{k+1}{2}} + \frac{2\mu_k^*(w)+w^2 D_k^*(w)-(l_k^{\rm uc})^2}{4} \label{ldf:lklm1_1.5}.
\end{split}
\end{equation}
To simplify this expression further we evaluate the chemical potential $\mu^*_k(w)$ first by substituting the expression of the constrained density Eq.~\eqref{dens-s:klm1} in Eq.~\eqref{klm1 chem eq} and we get
\begin{equation}\label{aklm1:chem0}
\begin{split}
\mu_k^*(w) &= \frac{y^2}{2}  -\int_{-l_k(w)}^{w} dy' \,\, \frac{\rho^*_k\left(y', w\right)}{|y'-y|^k}\\
&=\frac{y^2}{2}  -A_k \int_{-l_k(w)}^{\bar{l}_k(w)} dy' \frac{\left(l_k(w)+y'\right)^{\frac{k+1}{2}} \left(\bar{l}_k(w)-y'\right)^{\frac{k+3}{2}}}{(w-y')|y'-y|^k} -\frac{D_k^*(w)}{(w-y)^{k}}.
\end{split}
\end{equation}
It turns out to be non-trivial to compute it for arbitrary value of $y$. However it can be calculated  for specific values $y=-l_k(w)$ and $y=\bar{l}_k(w)$. Here we compute it for $y=\bar{l}_k(w)$, which formally gives
\begin{equation}\label{aklm1:chem1}
\begin{split}
\mu_k^*(w) &=\frac{\bar{l}_k(w)^2}{2}  -A_k \int_{-l_k(w)}^{\bar{l}_k(w)} dy' \frac{\left(l_k(w)+y'\right)^{\frac{k+1}{2}} \left(\bar{l}_k(w)-y'\right)^{\frac{3-k}{2}}}{(w-y')} \\&-\frac{D_k^*(w)}{(w-\bar{l}_k(w))^{k}}.
\end{split}
\end{equation}
To perform the above integral, we use the variable transformation given in Eq.~\eqref{vart:klm1} which simplifies the equation to
\begin{equation}
\begin{split}
\mu_k^*(w) &=\tilde{L}_k(w)^2 \Bigg[\frac{(1-n_k(w))^2}{2}  -A_k \int_{0}^{1} dz \frac{\left(1-z\right)^{\frac{k+1}{2}} z^{\frac{3-k}{2}}}{(z+h_k(w)-1)} \\&-\frac{h_k(w)^{\frac{k+1}{2}} \left(h_k(w)-1\right)^{\frac{3-k}{2}}}{k(k+1)}\Bigg]\\
&=\tilde{L}_k(w)^2 \Bigg[\frac{(1-n_k(w))^2}{2}  -A_k \frac{B\left(\frac{5-k}{2},\frac{k+3}{2}\right) ~_2F_1\left(\frac{5-k}{2},1,4,\frac{1}{1-h_k(w)}\right)}{h_k(w)-1} \\&-\frac{h_k(w)^{\frac{k+1}{2}} \left(h_k(w)-1\right)^{\frac{3-k}{2}}}{k(k+1)}\Bigg].
\end{split}
\end{equation}
Further using the hypergeometric identity in Eq.~\eqref{identity:hype} and Eq.~\eqref{qkwlm1}, one gets the following simplified equation
\begin{equation}\label{klm1 chem eq-1}
\mu_k^*(w) = \tilde{L}_k(w)^2 \left(\frac{1}{8 k}-\frac{h_k(w)-1}{2 k (k+1)}-\frac{(h_k(w)-1)^2 (k+2)}{2 k (k+1)^2}\right).
\end{equation}
To verify the validity of this expression we numerically perform the integral in Eq.~\eqref{aklm1:chem0} and compare it with Eq.~\eqref{klm1 chem eq-1} in Fig.~\ref{fig:klm1chem}.
The left LDF given in Eq.~\eqref{ldf:lklm1_1.5}, can be further simplified by using the expression of the chemical potential from Eq.~\eqref{klm1 chem eq-1} and representing the integrals in terms of hypergeometric functions,
\begin{equation}\label{aklm1:in}
\begin{split}
A_k \int_{0}^1& dz\, \frac{\left(1-n_k(w)-z\right)^2}{4}  \frac{z^{\frac{k+3}{2}} \left(1-z\right)^{\frac{k+1}{2}}}{z+h_k(w)-1}\\&= \frac{A_k}{4} \Bigg[n_k(w)^2\int_{0}^1 dz\,  \frac{z^{\frac{k+3}{2}} \left(1-z\right)^{\frac{k+1}{2}}}{z+h_k(w)-1}-2n_k(w)\int_{0}^1 dz\,  \frac{z^{\frac{k+3}{2}} \left(1-z\right)^{\frac{k+3}{2}}}{z+h_k(w)-1}\\&+\int_{0}^1 dz\,   \frac{z^{\frac{k+3}{2}} \left(1-z\right)^{\frac{k+5}{2}}}{z+h_k(w)-1}\Bigg] \\
&=\frac{A_k}{4(h_k(w)-1)} \Bigg[n_k(w)^2 B\left(\frac{k+3}{2},\frac{k+5}{2}\right)  ~_2F_1[\frac{k+5}{2},1,k+4, \frac{1}{1-h_k(w)}]\\&-2n_k(w)B\left(\frac{k+5}{2},\frac{k+5}{2}\right)~_2F_1[\frac{k+5}{2},1,k+5, \frac{1}{1-h_k(w)}]\\&+B\left(\frac{k+5}{2},\frac{k+7}{2}\right) ~_2F_1[\frac{k+5}{2},1,k+6, \frac{1}{1-h_k(w)}]\Bigg].
\end{split}
\end{equation}
Using the hypergeometric identity given in Eq.~\eqref{identity:hype} in Eq.~\eqref{aklm1:in} and simplifying, we get
\begin{equation}\label{aklm1:in1}
\begin{split}
A_k \int_{0}^1& dz\, \frac{\left(1-n_k(w)-z\right)^2}{4}  \frac{z^{\frac{k+3}{2}} \left(1-z\right)^{\frac{k+1}{2}}}{z+h_k(w)-1} \\&= \frac{A_k}{4} B\left(\frac{3 + k}{2}, \frac{3 + k}{2}\right) \\& \times ~_2F_1[1, -(k+2), -\frac{k+1}{2}, 1-h_k(w)] \Bigg[   n_k(w)^2  \\&-n_k(w) \frac{~_2F_1[1, -(k+3), -\frac{k+1}{2}, 1-h_k(w)]}{~_2F_1[1, -(k+2), -\frac{k+1}{2}, 1-h_k(w)]}   \\&+\frac{k+5}{4(k+4)} \frac{~_2F_1[1, -(k+4), -\frac{k+1}{2}, 1-h_k(w)]}{~_2F_1[1, -(k+2), -\frac{k+1}{2}, 1-h_k(w)]}\Bigg] \\&- \frac{h_k(w)^{\frac{k+1}{2}} (h_k(w)-1)^{\frac{k+3}{2}}}{4k(k+1)}\left(\frac{2(k+2)h_k(w) - (k+3)}{2(k+1)}\right)^2.
\end{split}
\end{equation}
This Eq.~\eqref{aklm1:in1} can be expressed further as
\begin{equation}\label{aklm1:in2}
\begin{split}
A_k \int_{0}^1& dz\, \frac{\left(1-n_k(w)-z\right)^2}{4}  \frac{z^{\frac{k+3}{2}} \left(1-z\right)^{\frac{k+1}{2}}}{z+h_k(w)-1} \\&= \frac{1}{4 \left(\tilde{L}_k(w)\right)^{k+2}} \Bigg[  n_k(w)^2  -n_k(w) \frac{~_2F_1[1, -(k+3), -\frac{k+1}{2}, 1-h_k(w)]}{~_2F_1[1, -(k+2), -\frac{k+1}{2}, 1-h_k(w)]}   \\&+\frac{k+5}{4(k+4)} \frac{~_2F_1[1, -(k+4), -\frac{k+1}{2}, 1-h_k(w)]}{~_2F_1[1, -(k+2), -\frac{k+1}{2}, 1-h_k(w)]}\Bigg] \\&- \frac{h_k(w)^{\frac{k+1}{2}} (h_k(w)-1)^{\frac{k+3}{2}}}{4k(k+1)}\left(\frac{2(k+2)h_k(w) - (k+3)}{2(k+1)}\right)^2.
\end{split}
\end{equation}
where we used Eq.~\eqref{aklm1:nom2}. We then use this expression along with the simplified expression of the chemical potential given in Eq.~\eqref{klm1 chem eq-1} and the normalization condition from Eq.~\eqref{aklm1:nom2} to get the final expression for left LDF given in Eq.~\eqref{ldf:lklm1_2}.

\section{Importance sampling method}
\label{apndims}
\begin{figure}[b]
    \centering
    \includegraphics[scale=0.6]{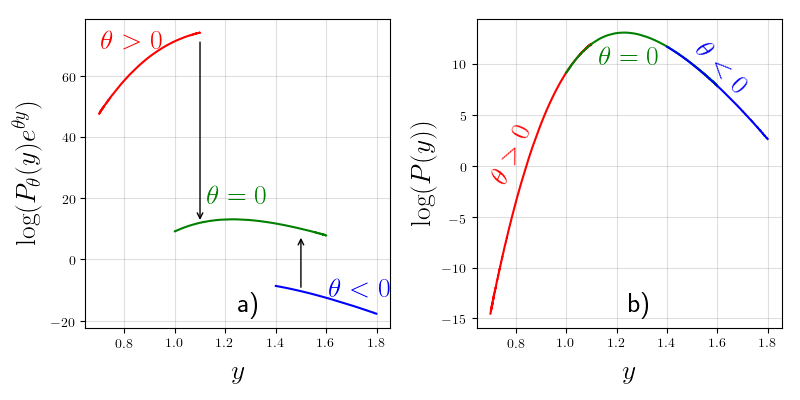}
    \caption{(a) We join the $P_{\theta}(y_{\max}) e^{\theta y_{\max}}$ for different values of $\theta$ by multiplying them by the appropriate normalization constant considering there is a small overlap in the argument of the distribution for two successive vaules of $\theta$. (b) This stitching procedure gives the distribution of $y_{\max}$ in the unbaised problem for rare fluctuations.}
    \label{fig:ims}
\end{figure}

Using a conventional Markov chain Monte-Carlo simulation using the Metroplis-Hashting algorithm, we can explore the probabilities of the order $n_{mc}^{-1}$, where $n_{mc} \equiv$ number of data points. However to compute the probability of the extremely rare events say of order $10^{-20}$ is not feasible to use this algorithm. Hence, we use importance sampling method~\cite{hartmann2011large, hartmann2018distribution, hartmann2018high} which is described here.

Probability of the position of the edge particle for a given ensemble is
\begin{equation}\label{al-1}
	P(y) = \sum_{\{y_i\}} Q[\{y_i\}] \delta(y-y_{\max}),
\end{equation}
where $Q[\{y_i\}]$ is the probability of a configuration $\{y_i\}$ and $y_{\max}$ is the position of the rightmost particle. To sample atypical values of $y_{\max}$ one needs to bias the sampling porcedure in the simulation. We run a Markov chain Monte-Carlo simulation using a biased Metroplis-Hastings Algorithm given by the following weight for the jump from configuration $\{y_i\}$ to $\{y'_i\}$:
\begin{equation}\label{acceptance}
	A(\{y_i\},\{y'_i\}) = \min \left(1, \frac{Q[\{y'_i\}]}{Q[\{y_i\}]}e^{(-\theta (y'_{\max}-y_{\max}))}\right).
\end{equation}
More precisely $A(\{y_i\}, \{y'_i\})$ is the probability of acceptance of the new configuration ($\{y'_i\}$) given a configuration ($\{y_i\}$). Note that $\theta>0$ biases the ensemble to have smaller $y_{\max}$ and $\theta<0$ does the opposite. The stationary distribution of this Markov chain is given by $Q_{\theta}[\{y_i\}] = Q[\{y_i\}]e^{-\theta y_{\max}}$. 

Now using this algorithm one can numerically compute the PDF of $y_{\max}$ in the biased simulation which we denote by $P_{\theta}(y)$. We have
\begin{equation}
\begin{split}
P_{\theta}(y) &= \sum_{\{y_i\}} Q_{\theta}[\{y_i\}] \delta(y-y_{\max}),\\
			  &= \sum_{\{y_i\}} Q[\{y_i\}]e^{-\theta y_{\max}} \delta(y-y_{\max}),\\
	 		&= e^{-\theta y} \sum_{\{y_i\}} Q[\{y_i\}] \delta(y-y_{\max}),
\end{split}
\end{equation}
which from Eq.~\eqref{al-1} gives
\begin{equation}\label{ims:ptp}
	P(y) = \sum_{\{y_i\}} Q[\{y_i\}] \delta(y-y_{\max}) = P_{\theta}(y) e^{\theta y}. 
\end{equation}
Using this relation Eq.~\eqref{ims:ptp}, we can compute the PDF [$P(y_{\max} = y)$] of extremely large fluctuations of $y_{\max}$ which can be obtained by considering larger absolute values of $\theta$. However for any $\theta$ the width of the distribution obtained numerically is finite. Hence to compute the PDF which also includes the rare fluctuations one has to simulate $P_{\theta}(y)$ for many values of $\theta$ and these different biased PDF's are glued together by multiplying by the appropriate normalization prefactor to $P_{\theta}(y)$ such that $\theta \neq 0$ connects smoothly with $\theta=0$. A schematic cartoon of this procedure is shown in Fig.~\ref{fig:ims}. %In principle one can get $P(y)$ of the order of $10^{-1000}$ by considering greater and greater absolute value of $\theta$.

\begin{center}
\line(1,0){250}
\end{center}
%\end{appendices}

\clearpage
\section*{References}
%\bibliographystyle{iopart-num}
%\bibliography{cita}

%
%\section*{References}
%
%% 
%% 

%%%%%%%%%%%%%%%%%%%%%%%%%%%%%%%%%%%%%%%%%
% \bibliographystyle{unsrt}
% \bibliography{cita}
%%%%%%%%%%%%%%%%%%%%%%%%%%%%%%%%%%%%%%%%%
\end{document}